\documentclass[paper=a4, fontsize=11pt]{scrartcl} % A4 paper and 11pt font size

\usepackage[T1]{fontenc} % Use 8-bit encoding that has 256 glyphs
\usepackage{fourier} % Use the Adobe Utopia font for the document - comment this line to return to the LaTeX default
\usepackage[english]{babel} % English language/hyphenation

\usepackage{mathtools}
\usepackage{amsmath,amssymb,euscript,appendix,marvosym,bbm,enumitem}
\usepackage{arydshln}
\usepackage{todonotes}
\usepackage{graphicx}
\usepackage{float}
\usepackage[margin=1cm,font=small]{caption}
\usepackage[margin=.2cm,font=footnotesize]{subcaption}
\usepackage[normalem]{ulem}
\usepackage{mdframed}
\usepackage{bm}
\usepackage{comment}
\usepackage{wasysym}
\usepackage{enumitem}
\usepackage{xcolor}
\usepackage[bottom]{footmisc}
\usepackage{stmaryrd}
\usepackage{xfrac}
\usepackage{cite}
\usepackage{booktabs}

\usepackage[linkcolor=blue,colorlinks=true]{hyperref}
\usepackage{cleveref}
\hypersetup{
    colorlinks,
    linkcolor={blue},
    citecolor={blue},
    urlcolor={purple}
}
 
 \usepackage{geometry}
\usepackage{braket}

\DeclareFontFamily{U}{mathx}{\hyphenchar\font45}
\DeclareFontShape{U}{mathx}{m}{n}{<-> mathx10}{}
\DeclareSymbolFont{mathx}{U}{mathx}{m}{n}
\DeclareMathAccent{\widebar}{0}{mathx}{"73}

\DeclareMathAlphabet{\mathdsl}{U}{bbm}{m}{sl}

\usepackage{sectsty} % Allows customizing section commands
\usepackage{fancyhdr} % Custom headers and footers
\numberwithin{equation}{section} % Number equations within sections (i.e. 1.1, 1.2, 2.1, 2.2 instead of 1, 2, 3, 4)
\numberwithin{figure}{section} % Number figures within sections (i.e. 1.1, 1.2, 2.1, 2.2 instead of 1, 2, 3, 4)
\numberwithin{table}{section} % Number tables within sections (i.e. 1.1, 1.2, 2.1, 2.2 instead of 1, 2, 3, 4)

\newcommand{\horrule}[1]{\rule{\linewidth}{#1}} % Create horizontal rule command with 1 argument of height

\title{	
\normalfont \normalsize 
\textsc{} \\ [25pt] % Your university, school and/or department name(s)
\horrule{0.5pt} \\[0.4cm] % Thin top horizontal rule
\huge 
 Particle production in a light-cone gauge fixed Jordanian deformation of $AdS_5\times S^5$    \\% The assignment title
\horrule{2pt} \\[0.5cm] % Thick bottom horizontal rule
}

\author{Riccardo Borsato$^\circ$ and Sibylle Driezen$^\diamond$} % Your name

\date{} % Today's date or a custom date  \normalsize\today

\begin{document}

\maketitle % Print the title

\begin{center}
{
\small {\textit{ 
$\circ$ Instituto Galego de F\'isica de Altas Enerx\'ias (IGFAE) and
Departamento de F\'\i sica de Part\'\i culas,\\[7pt]
Universidade de  Santiago de Compostela, Spain\\
\vspace{12pt}
$\diamond$ Institut f\"ur Theoretische Physik, ETH Z\"urich,\\[7pt]
Wolfgang-Pauli-Strasse 27, 8093 Z\"urich, Switzerland\\
\vspace{12pt}
\texttt{riccardo.borsato@usc.es, sdriezen@phys.ethz.ch}}}}\\

\vspace{100pt}

\textbf{Abstract}
\end{center}
\noindent
We consider a string on a Jordanian deformation of the $AdS_5\times S^5$ spacetime. This model  belongs to the larger class of Homogeneous Yang-Baxter deformations, which preserve classical integrability in the sense that one can construct an explicit Lax connection. To study the scattering of bosonic worldsheet excitations, we fix light-cone gauge and expand around  a pointlike classical solution that reduces to the BMN vacuum in the undeformed limit. Our analysis shows that the light-cone gauge-fixed Hamiltonian, under a perturbative field expansion, includes cubic terms that give rise to  non-trivial cubic processes for physical particles. We discuss this unexpected result  in relation to the property of Lax integrability of the sigma-model.
\thispagestyle{empty}

\pagebreak

\setcounter{page}{1}
\newcounter{nameOfYourChoice}

\tableofcontents

\section{Introduction}

The study of integrable structures   in the context of the AdS/CFT correspondence has provided remarkable results on the dynamics of certain gauge and string theories at finite values of the coupling in the planar limit. The classic example is the discrete integrable system underlying type IIB superstrings on $AdS_5 \times S^5$ and its dual, ${\cal N} = 4$ Super Yang-Mills theory. Integrable methods then enabled the exact computation of several observables for these models, cf.~\cite{Beisert:2010jr} for a review.

\vspace{12pt}

A natural progression of this framework involves the study of integrable deformations of the $AdS_5 \times S^5$ superstring. Among these, the Homogeneous Yang-Baxter (HYB) deformations  \cite{Klimcik:2008eq,Delduc:2013fga,Delduc:2013qra,Kawaguchi:2014qwa,vanTongeren:2015soa} are particularly interesting since, as first pointed out in~\cite{vanTongeren:2015uha}, the modification of the integrable structure of $AdS_5\times S^5 / {\cal N}=4$ SYM seems to be elegantly related to Drinfel’d twists \cite{drinfeld1983constant}, see also~\cite{Giaquinto:1994jx,Kulish2009}. 
This is, in fact, particularly well-understood for the restricted subclass of ``diagonal’’ TsT (T-duality-shift-T-duality) transformations, also known as ``diagonal’’ abelian HYB deformations,\footnote{Here, the adjective ``diagonal’’ refers to twists or TsT transformations over Cartan isometries. The general reformulation of abelian HYB deformations in terms of TsT transformations was established in \cite{Osten:2016dvf}.} where the twist of the integrable structure is established through a Drinfel’d-Reshetikhin twist~\cite{Reshetikhin:1990ep}. This allowed an efficient application of the integrability methods, see e.g.~ \cite{Beisert:2005if,Ahn:2010ws,deLeeuw:2012hp,Kazakov:2018ugh,vanTongeren:2021jhh}. However, beyond this special subclass, a uniform picture of all HYB deformations---including ``non-diagonal’’ TsT transformations
and Jordanian deformations---is lacking.\footnote{Notably, an initial study of the spectrum for a specific non-diagonal TsT-deformed model, corresponding to a dipole deformation of ${\cal N}=4$ SYM, was done in~\cite{Guica:2017mtd}. This work utilised the associated Drinfel'd twisted spin chain.}
Below, let us  elaborate on the several reasons why HYB deformations merit attention in this context: 

$\bullet$ At the level of the worldsheet sigma-model, HYB deformations are realised through an antisymmetric ``$r$-matrix’’ which is a solution of the Classical Yang-Baxter equation (CYBE). Importantly, this property ensures that HYB deformations preserve classical integrability, through the existence of a Lax connection which is flat upon its equations of motion \cite{Klimcik:2008eq,Delduc:2013fga,Delduc:2013qra}. Antisymmetric solutions of the CYBE are in fact known to be in one-to-one correspondence with Drinfel’d twists that are continuously connected to the identity \cite{drinfeld1983constant}. 

$\bullet$ The deformation of the Hamiltonian and Poisson structure can be understood as a \textit{non-local} canonical transformation \cite{Vicedo:2015pna}.
The non-locality warrants that the HYB deformation is non-trivial and that, at least classically and on-shell, its effect can be completely mapped to a deformation of the boundary conditions of the sigma-model fields~\cite{Frolov:2005dj,vanTongeren:2018vpb,Borsato:2021fuy}. This fact mimics the understanding of Drinfel’d twists in the context of discrete integrable models like spin chains~\cite{drinfeld1983constant}. 

$\bullet$ The  background fields of the deformed $AdS_5\times S^5$  satisfy the supergravity equations of motion if the $r$-matrix solution is unimodular with respect to the original algebra $\mathfrak{g}$ of isometries 
\cite{Borsato:2016ose}. Unimodular HYB deformations are thus well-suited for interpretation within AdS/CFT. For $\mathfrak{g} = \mathfrak{psu}(2,2|4)$, relevant to $AdS_5 \times S^5$, a (possibly incomplete) classification comes from considering all possible even-dimensional abelian subalgebras of $\mathfrak{g}$, together with the non-abelian $r$-matrices listed in~\cite{Borsato:2016ose,Borsato:2022ubq}.

$\bullet$ In the dual gauge theory, HYB deformations are conjectured to modify the field product into a non-commutative star product through a Drinfel'd twist \cite{vanTongeren:2015uha}, see also~\cite{vanTongeren:2016eeb,Araujo:2017jkb}. The twist  can break both   internal and spacetime symmetries of the gauge theory. In  ${\cal N}=4$ SYM, for instance,  this results in  exactly marginal deformations (such as the $\beta$-deformation \cite{Leigh:1995ep,Lunin:2005jy}) 
or  the introduction of noncommutativity between spacetime coordinates \cite{Seiberg:1999vs,Bergman:2000cw}. 
Recently, significant progress has been made in formulating gauge-invariant Yang-Mills actions on noncommutative spacetimes, particularly for twists based on the Poincar\'e algebra, which can be understood as a subalgebra of $\mathfrak{g}=\mathfrak{psu}(2,2|4)$~\cite{Meier:2023kzt,Meier:2023lku}.

\vspace{12pt}

One of the key challenges in extending HYB-twisted gauge-gravity duality beyond the ``diagonal’’ abelian case  lies in the breaking of the Cartan subalgebra of isometries in the string background. This includes the light-cone isometries  used to fix uniform light-cone gauge in the undeformed worldsheet theory, which played a crucial role in the perturbative formulation of the integrable  worldsheet scattering \cite{Arutyunov:2004yx,Arutyunov:2005hd,Arutyunov:2006gs,Arutyunov:2009ga}. When these isometries are broken, one is forced to pick an alternative light-cone gauge. The consequences of doing this were explained in \cite{Borsato:2023oru}, see also \cite{Frolov:2019xzi,Hoare:2023zti} for related works. As mentioned above, an alternative perspective involves reformulating these deformations in terms of twisted boundary conditions for the string. This approach enabled the application of  integrable methods, such as the Classical Spectral Curve and its semiclassical quantisation,  to extract the semiclassical worldsheet spectrum for  certain non-diagonal TsT and Jordanian deformations \cite{Ouyang:2017yko,Borsato:2022drc,Driezen:2024mcn}. While the asymptotics of the Curve, which  encode the local charges, are  nontrivial, the Jordanian subclass is particularly well-defined, due to its polynomial asymptotics and  the diagonalisability of the twist appearing for the new boundary conditions. Despite this progress, a comprehensive understanding of the worldsheet dynamics of these models, particularly the scattering matrix, remains incomplete.

\vspace{12pt}

In this paper, we will therefore focus on the specific Jordanian deformation of $AdS_5 \times S^5$ which was first constructed in~\cite{vanTongeren:2019dlq}, and then studied in \cite{Borsato:2022drc} with spectral curve methods. The main aim of the present work is to make progress on the  perturbative formulation of its worldsheet scattering. After implementing a light-cone gauge using a classical pointlike solution, 
we will study the tree-level scattering of its bosonic worldsheet excitations. Already at tree-level, we obtain a surprising result: Due to a cubic Hamiltonian, the Jordanian deformation appears to exhibit particle production in the light-cone gauge, with cubic processes that are  non-trivial when going on-shell and relaxing the level-matching condition (i.e.~when the total momentum is not zero). This feature clashes with the usual axioms of integrable S-matrices, and it therefore challenges the naive expectations coming from the Lax integrability of the string sigma-model. We will discuss its possible explanations and implications.

\vspace{12pt}

The article is organised as follows: In section \ref{s:preliminaries}, we introduce the necessary ingredients for the study of the bosonic worldsheet scattering of this Jordanian sigma-model, i.e.~the deformed background geometry as well as the classical pointlike solution of the equations of motion. 
We use this solution to fix the (alternative) light-cone gauge in section \ref{s:lcgf}, after which we present the perturbative tree-level Hamiltonian until cubic order in the fields in section \ref{s:pert-ham}. In section  \ref{s:quant-ham}, we introduce the creation and annihilation operators of the bosonic excitations, as well as their deformed  dispersion relations, which are generally non-relativistic. We then show that on-shell we have non-vanishing processes with a three-point vertex, implying the production of physical particles.
We end in section \ref{s:discussion} with a final discussion about our results. 
For the interested reader, we  added an ancillary \textsf{Mathematica} file to the \textsf{arXiv} submission where we summarise the main calculations needed to verify the presence of particle production at tree-level.

\section{Preliminaries} \label{s:preliminaries}
%s

\subsection{The background}
The Homogeneous Yang-Baxter deformation that we study in this paper is of Jordanian type and  is defined by the following antisymmetric $r$-matrix solution of the CYBE
\begin{equation} \label{eq:r-matrix}
    r = \mathsf{h} \wedge \mathsf{e}, \qquad \text{with} \qquad \mathsf{h} = \frac{D - J_{03}}{2} , \qquad \mathsf{e} = \sqrt{2} \left(p_0 + p_3\right),
\end{equation}
where $D$ is the dilatation generator, $J_{ij}$ the Lorentz generators, and $p_i$ the translation generators of the conformal subalgebra $\mathfrak{so}(2,4)$ of $\mathfrak{psu}(2,2|4)$. We refer to~\cite{Borsato:2022ubq} for our conventions on the (super)algebra $\mathfrak{psu}(2,2|4)$ and a possible matrix realisation.\footnote{Compared to  previous works \cite{Borsato:2022ubq,Borsato:2022drc}, we note that we have rescaled the generator $\mathsf{e}$, which amounts to rescaling the deformation parameter $\eta$.}
Given a Yang-Baxter deformation, the corresponding background is a solution of the type IIB supergravity equations when $r$ is unimodular \cite{Borsato:2016ose}, which means that the above $r$-matrix should be extended to include certain supercharges of $\mathfrak{psu}(2,2|4)$, see for example \cite{Borsato:2022ubq,vanTongeren:2019dlq}. However, this extension is irrelevant in the context of this paper, because here we will only study the bosonic truncation of the sigma-model. 

\vspace{12pt}

The  deformed background metric and Kalb-Ramond two-form in (polar) Poincar\'e coordinates are \cite{Kawaguchi:2014fca} (see~\cite{Borsato:2022drc} for our conventions)
\begin{equation}
    \begin{aligned}
        ds^2 &= \frac{dz^2 + d\rho^2 + \rho^2 d\theta^2 - 2 dx^+ dx^-}{z^2}-\frac{\eta^2 (z^2+\rho^2)dx^{-2}}{ z^6} + ds^2_{S^5} ,  \\
        B&= \eta \frac{\rho d\rho \wedge dx^-}{z^4} - \eta d \left( \frac{dx^-}{2 z^2} \right) ,
    \end{aligned}
\end{equation}
where $ds^2_{S^5}$ is the metric of the round $S^5$ sphere. As usual, it is convenient to  parameterise the latter  by global  Stereographic-Hopf coordinates $(\phi, y_i~;~i=1,\ldots,4)$ as done e.g.~in~\cite{Arutyunov:2009ga}
\begin{equation}
    ds^2_{S^5} = \left( \frac{1-\frac{y^2}{4}}{1+\frac{y^2}{4}} \right)^2 d\phi^2  + \frac{dy_i dy_i}{\left( 1+\frac{y^2}{4}\right)^2} , \qquad\quad y^2 = y_iy_i , 
\end{equation}
with $\phi \in [0,2\pi]$ parameterising a big circle in $S^5$. 
As shown in \cite{Borsato:2022drc}, a global coordinate system  for the deformed $AdS$ space can be obtained by the transformation\footnote{See also \cite{Blau:2009gd} for the introduction of these global coordinates for Schr\"odinger spacetimes.}
\begin{equation}
    \begin{gathered}
        x^+  = V + \frac{1}{2} \left( Z^2 + x_2^2 + x_3^2 \right) \tan T , \qquad z = \frac{Z}{\cos T} , \qquad \rho = \frac{\sqrt{x_2^2 + x_3^2}}{\cos T} , \\
        \theta = \arctan\left(\frac{x_2}{x_3}\right)  , \qquad x^- = \tan T ,
    \end{gathered}
\end{equation}
so that now the deformed background fields read
\begin{equation} \label{eq:G-B-Sch}
    \begin{alignedat}{2}
        ds^2 &= G_{MN} dX^M dX^N = \frac{dZ^2 + dx_2^2 + dx_3^2 - 2dT dV}{Z^2}  - \frac{(Z^4+\eta^2)(Z^2+x_2^2+x_3^2) }{Z^6}dT^2 + ds^2_{S^5} ,  \\
B &= \frac{1}{2} B_{MN} dX^M \wedge dX^N =  \eta \frac{x_2 dx_2 \wedge dT+x_3 dx_3 \wedge dT}{Z^4} - \eta d \left( \frac{    dT}{2Z^2} \right)  .
    \end{alignedat}
\end{equation}
We will call this the global coordinate system, with coordinates $X^M=(T,V,x_2,x_3,Z, \phi, y_i)$. 
The background is further supported in type IIB supergravity  by a non-trivial $F_3$ and $F_5$ RR-flux whilst the dilaton is constant $\Phi=\Phi_0$.\footnote{The explicit expressions of the deformed IIB supergravity background can be found in Eq.~(7) of \cite{Driezen:2024mcn} after setting $a=0$ therein and transforming $( x_2=P \cos\Theta, x3 = P \sin\Theta )$. This background was first constructed in~\cite{Kawaguchi:2014fca}, see also~\cite{Matsumoto:2014ubv,vanTongeren:2019dlq}.} Note that we can remove the total derivative, $d (Z^{-2}dT) $, in the background B-field, and we will do so in the remainder of this paper. 

\vspace{12pt}

The (manifest) residual isometry subalgebra of this target space 
corresponds to those generators $\mathsf{T}_A\in \mathfrak{psu}(2,2|4)$ whose adjoint action commutes with the action of the $r$-matrix. In our case, we  have five residual isometries in $\mathfrak{so}(2,4)$, given by 
\begin{equation}
    \mathfrak{t}_{\mathfrak{a}} = \mathrm{span} \left( D+J_{03}, k_0+k_3, p_0, p_3, J_{12} \right) \cong \mathfrak{sl}(2,R) \oplus \mathfrak{u}(1)^2. 
% \subset \mathfrak{so}(2,4) .
\end{equation}
When including the unimodular extension in the $r$-matrix, only an $\mathfrak{su}(3)\oplus \mathfrak{u}(1)$ subalgebra of $\mathfrak{so}(6)$ is preserved, 
as well as 12 supercharges \cite{Borsato:2022ubq}. However, in the bosonic sigma model considered in this paper, we will still have the full $SO(6)$ symmetry of the $S^5$.
The Cartan subalgebra of the residual symmetry algebra $\mathfrak{t}_{\mathfrak{a}}$ is three-dimensional and generated by \cite{Borsato:2022drc}
\begin{equation}
   \mathsf{H}_T = \frac{1}{2} \left(p_0 - k_0 - p_3 - k_3 \right) , \qquad \mathsf{H}_V = p_0 + p_3 , \qquad \mathsf{H}_\Theta = J_{12} .
\end{equation}
These are manifestly realised in the global coordinate system \eqref{eq:G-B-Sch} as shifts of $T,V$ and $\Theta=\arctan (x_2/x_3)$ respectively. Knowing that $\mathsf{H}_T$ is a timelike generator, the coordinate $T$ has the interpretation of global time. Importantly, it is not the usual BMN time direction, see~\cite{Borsato:2022drc} for comments on this.

\subsection{The classical solution } \label{s:class-sol}
In the next sections, we will study the fluctuations around a classical solution of the sigma-model equations of motion. To do so, we find it convenient to introduce a new set of coordinates, related to the previous ones through the following transformations:
\begin{equation} \label{eq:coord-trans-lin} 
    T = \frac{t-v}{\sqrt{1-\eta^2}},\qquad
    V= \frac{v-\eta^2 t}{\sqrt{1-\eta^2}},\qquad
    Z = 1+z ,
\end{equation}
with $t$ the new  time coordinate.
Due to  the linearity of this transformation,  the deformed  background remains invariant under   independent shifts of the  $t$  and $v$ coordinates. Lastly, we define the light-cone coordinates\footnote{Although we use the same notation  $(x^\pm,z)$ as in the Poincar\'e coordinate system, these coordinates differ from those used in that context -- we trust that no confusion  should arise. }
\begin{equation} \label{eq:coord-trans-lc} 
    x^+ = (1-a)t+a\phi, \qquad x^- =\phi-t ,
\end{equation}
with $a$ an arbitrary real number, that is customary to introduce as a gauge parameter for light-cone gauge fixing. 
It is straightforward to show that a  possible configuration solving the equations of motion is to set all fields $(x^-,v,z,x_2,x_3,y_i)$  to zero, while 
\begin{equation}
    \bar x^+=\tau.
\end{equation}
In this equation, as well as in the rest of the paper, a bar denotes quantities evaluated on the chosen classical solution. To verify that the configuration above  is indeed a solution, we start from the general form of the sigma-model equations of motion
\begin{equation} \label{eq:string-eom}
    \partial_\alpha \left( \gamma^{\alpha\beta} \partial_\beta X^M \right) + \Gamma_{NK}^{(-)\alpha\beta M} \partial_\alpha X^N \partial_\beta X^K = 0 ,
\end{equation}
where $\Gamma_{NK}^{(-)\alpha\beta M} =\gamma^{\alpha\beta} \Gamma^M_{NK} - \frac{1}{2} \epsilon^{\alpha\beta} G^{MP}H_{PNK}$. Here $\Gamma^M_{NK}$ are the Christoffel symbols, $H_{MNP}$ is the field strength of the $B$-field, and $\gamma^{\alpha\beta}=\sqrt{|h|}h^{\alpha\beta}$ with $h_{\alpha\beta}$ the worldsheet metric, so that $\det\gamma=-1$. When $\gamma^{\alpha\beta}$ is taken to be constant and the solution is point-like as above (i.e.,~it  depends only on $\tau$ and not on $\sigma$), the equations of motion reduce to the standard geodesic equations for a particle in a curved background
\begin{equation}
    \Ddot{X}^M+\Gamma^M_{NK}\dot X^N\dot X^K=0,
\end{equation}
where the dot denotes differentiation with respect to $\tau$.
Given that our fields are at most linear in $\tau$, and that in the classical configuration the only non-trivial derivative is $\dot{\bar x}^+=1$, the geodesic equation simply implies
\begin{equation}
  \bar  \Gamma^M_{++}=0,\qquad \forall M.
\end{equation}
This condition can be directly verified by computing  the Christoffel symbols in this coordinate system.

At the same time, we need to ensure that the classical configuration also satisfies the Virasoro constraints. These constraints arise from demanding that the worldsheet stress-energy tensor 
\begin{equation}
    T_{\alpha\beta} = \partial_\alpha X^M G_{MN} \partial_\beta X^N - \frac{1}{2} \gamma_{\alpha\beta} \gamma^{\gamma\delta} \partial_\gamma X^M G_{MN} \partial_\delta X^N ,
\end{equation}
vanishes identically.
Evaluating this  expression on the classical solution yields
\begin{equation}
     \bar   T_{00}=\bar G_{++}\left(1-\frac{1}{2}  \gamma_{00}  \gamma^{00}\right),\qquad
        \bar T_{01}=-\frac{1}{2}\bar G_{++}  \gamma_{01}  \gamma^{00},\qquad
        \bar T_{11}=-\frac{1}{2}\bar G_{++}  \gamma_{11}  \gamma^{00}.
\end{equation}
Using the fact that $\bar G_{++}=0$, we can conclude that the classical configuration satisfies the Virasoro constraints, as required.
It is worth noting that  in the $\eta\to 0$ limit, this classical configuration reduces to the standard BMN point-like solution typically considered in the case of $AdS_5\times S^5$, see~\cite{Borsato:2022drc} for comments on this.

\vspace{12pt}

To obtain a more convenient expression for the Hamiltonian density of the gauge-fixed model (that will be derived in the next sections) we actually prefer to implement yet another  coordinate redefinition in target space. Specifically, we define
\begin{equation} \label{eq:coord-trans-quadcub} 
    x^- = \tilde x^--\tilde v\tilde z(1+2\tilde z)+\frac23 \tilde v^3,\qquad
    v= \tilde v+2\tilde v\tilde z,\qquad
    z=\tilde z+\frac{\tilde z^2-2\tilde v^2- \tilde x_2^2- \tilde x_3^2}{2},\qquad
    x_a=\tilde x_a+\tilde x_a\tilde z ,
\end{equation}
where the index $a$ labels the $x_2$ and $x_3$ fields.
In these new coordinates, the  background will  appear more complicated, but the perturbative Hamiltonian density of the gauge-fixed model simplifies, at least to the order   relevant for our discussion. Note in fact that we shift $x^-$ by terms that are at most cubic in the other coordinates, and $v,z$ and $x_a$ by expressions that are at most quadratic. Since our perturbative analysis of the Hamiltonian will  stop at cubic order,  higher-order coordinate redefinitions would have no impact in our considerations.
For readability, we will drop the tildes from the new coordinates after applying these redefinitions. 

Importantly, note that this redefinition does not change the form of the classical configuration; i.e.~the classical solution remains $\bar x^+=\tau$ with all other fields set to zero.
For an interpretation of the above coordinate transformations at the level of the gauge-fixed sigma model, see for example~\cite{Borsato:2023oru}.

\section{Light-cone gauge-fixing} \label{s:lcgf}
Having identified a viable classical configuration, we now study the fluctuations around it by imposing the uniform light-cone gauge~\cite{Kruczenski:2004kw,Kruczenski:2004cn,Arutyunov:2004yx}. This procedure of light-cone gauge-fixing is well-established (see e.g.~\cite{Arutyunov:2009ga}), so here we review only the essential aspects. We start from the classical sigma-model action
\begin{equation}\label{eq:polyakov}
    S = -\frac{g}{2}\int d\tau d\sigma\ 
    \left(\gamma^{\alpha\beta}G_{MN}-\epsilon^{\alpha\beta}B_{MN}\right)\partial_\alpha X^M\partial_\beta X^N,
\end{equation}
where $g$ represents the string tension, $X^M = \{x^+,x^-,v,x_a,z,y_i\}$, and we take the convention $\epsilon^{\tau\sigma}=-1$.
Then, we define the conjugate momenta via a  Legendre transformation
\begin{equation}
p_M = \frac{\delta S}{\delta \dot X^M} 
= -g\gamma^{0\beta}\partial_\beta X^NG_{MN}+gX'{}^NB_{MN}.   
\end{equation} 
Notice that, on the classical configuration, the momenta simplify to
\begin{equation}
\bar p_M=-g\bar \gamma^{00}\bar G_{M+}.
\end{equation} 
A key advantage of using the  coordinate system introduced in section \ref{s:class-sol}  is  that, after taking $\bar \gamma^{00}=-1/g$ on the classical configuration, the classical momenta take  the convenient constant  values
\begin{equation}
    \bar p_+=0,\qquad \bar p_-=1, \qquad \bar p_\mu=0,
\end{equation}
where the index $\mu$ denotes  directions other than $M=+$ or $M=-$. These directions labelled by $\mu$ are referred to as ``transverse'', as they are not gauge-fixed and represent the physical degrees of freedom.

A standard calculation shows that the action~\eqref{eq:polyakov}  can be rewritten in first-order form as
\begin{equation}\label{eq:action-first-order}
    S = \int d^2\sigma  \left( p_M \dot{X}^M + \frac{\gamma^{01}}{\gamma^{00}} C_1 + \frac{1}{2g \gamma^{00}} C_2 \right),
\end{equation}
where the expressions
\begin{equation}
\begin{aligned}
    C_1&=p_MX'{}^M,\\
        C_2& = G^{MN}p_Mp_N+g^2 G_{MN}X'{}^MX'{}^N-2g p_MG^{MN}B_{NQ}X'{}^Q+g^2G^{MN}B_{MP}B_{NQ}X'{}^PX'{}^Q,
        \end{aligned}
\end{equation}
implement the Virasoro constraints in the form $C_1=C_2=0$.
Then, considering the fluctuations $\hat x^M,\hat p_M$ of the fields around the classical configuration  $\bar x^M,\bar p_M$
\begin{equation}
    x^M = \bar x^M+\hat x^M,\qquad
    p_M = \bar p_M+\hat p_M,
\end{equation}
the procedure of light-cone gauge-fixing consists in imposing  that the fluctuations $\hat x^+$ and $\hat p_-$ are gauge-fixed to zero
\begin{equation}
    \hat x^+=0 \implies x^+=\tau,\qquad
    \hat p_- =0\implies p_-=1.
\end{equation}
Since we are setting to zero the fluctuations of the only two fields with non-vanishing classical values,  we can omit the hat notation for the fluctuations of the remaining fields, as we have {e.g.}~$x^\mu=\hat x^\mu$.

The procedure of uniform light-cone gauge-fixing provides an efficient way  to solve the Virasoro constraints. First, the condition $C_1=0$ is solved by
\begin{equation}
    x^-{}'=-p_\mu x^\mu{}'.
\end{equation}
The second condition $C_2=0$ is a quadratic equation in $p_+$. Introducing indices $m,n=-,\mu$ (i.e.~all except $+$),  the equation can be written as as $C_2=Ap_+^2+Bp_++C=0$, where\footnote{In all these expressions, note that $p_-$ should be replaced by  its classical value $p_-=\bar p_-=1$.}
\begin{equation}
    \begin{aligned}
        A&=G^{++},\\
        B&= 2G^{+m}p_m-2g G^{+M}B_{Mn}X'{}^n,\\
    C&= G^{mn}p_mp_n+g^2 G_{mn}X'{}^mX'{}^n-2g p_mG^{mN}B_{Nq}X'{}^q+g^2G^{MN}B_{Mp}B_{Nq}X'{}^pX'{}^q,
    \end{aligned}
\end{equation}
and we can solve it for $p_+$ as\footnote{We have chosen the sign of the square root as in~\cite{Arutyunov:2009ga}. This choice ensures that the Hamiltonian has a well-behaved perturbative expansion. Choosing the opposite sign would lead to a quadratic Hamiltonian that is not positive definite, and it would introduce various factors of $(1-2a)^{-1}$ that would diverge for the value $a=1/2$ of the gauge parameter.}
\begin{equation}
    p_+=\frac{-B+\sqrt{B^2-4AC}}{2A}.
\end{equation}
When the Virasoro constraints $C_1=C_2=0$ are imposed,  the action~\eqref{eq:action-first-order} simplifies to
\begin{equation}
    S = \int d^2\sigma \left( p_+ +\dot{x}^-+p_\mu \dot{x}^\mu  \right) = \int d^2\sigma \left( p_+ +p_\mu \dot{x}^\mu  \right),
\end{equation}
where in the second step we have dropped a total derivative. We recognise the action for the  fields $x^\mu,p_\mu$ with Hamiltonian density $\mathcal H=-p_+$, which is 
\begin{equation}\label{eq:fullH}
    \mathcal H=\frac{B-\sqrt{B^2-4AC}}{2A}.
\end{equation}
This result can then be interpreted as the Hamiltonian density for the ``transverse fields'' $x^\mu$ and $p_\mu$ only. The remaining fields are either gauged-fixed to their classical values ($x^+,p_-$) or  expressed in terms of $x^\mu,p_\mu$ through the Virasoro constraints ($x^-{}',p_+$).

\section{The perturbative Hamiltonian} \label{s:pert-ham}
The Hamiltonian density~\eqref{eq:fullH} derived from the expressions of background fields \eqref{eq:G-B-Sch} in the coordinate system obtained after the sequence of coordinate transformations \eqref{eq:coord-trans-lin}, \eqref{eq:coord-trans-lc} and \eqref{eq:coord-trans-quadcub} is quite complicated. To quantise it, it is convenient to implement a field expansion. This is  achieved in the so-called ``decompactification limit'' by rescaling the spatial coordinate $\sigma$ and the transverse fields as follows\footnote{We have verified that this scaling of transverse  fields, together with a scaling $x^+ \rightarrow \mu x^+$ and $x^- \rightarrow \mu^{-1}g^{-2} x^-$, leads to a well-defined plane-wave limit ($g\rightarrow\infty)$ and flat space limit ($g\rightarrow\infty$ and $\mu \rightarrow 0$).}
\begin{equation}
    \sigma\to g\sigma,\qquad
    x^\mu\to g^{-1}x^\mu,\qquad
    p_\mu\to g^{-1}p_\mu,
\end{equation}
and then sending $g\to \infty$. Under this scaling, the Hamiltonian density organises into an  expansion 
\begin{equation}
    \mathcal H = \mathcal H_2 + g^{-1}\mathcal H_3+ g^{-2}\mathcal H_4 + \cdots,
\end{equation}
where $\mathcal H_n$ represents the terms of degree $n$ in the  fields and  their derivatives.

\vspace{12pt}

\noindent \textbf{{The quadratic Hamiltonian}} --- At  quadratic order, the Hamiltonian density is\footnote{To write this result we removed the total derivative $\eta (1-\eta^2)^{\sfrac{-1}{2}}(x_2x'_2+x_3x'_3)$.}
\begin{equation} \label{eq:H2}
    \mathcal H_2 = \frac12\sum_\mu\left((p_\mu )^2+(x'_\mu)^2+m_\mu^2(x_\mu)^2\right)+p_zv-p_vz, 
\end{equation}
where 
\begin{equation}
    m_v^2=1,\qquad
    m_z^2=\frac{1+3\eta^2}{1-\eta^2},\qquad
    m_{x_a}^2=\frac{1+\eta^2}{1-\eta^2},\qquad
    m_{y_i}^2=1.
\end{equation}
Thus, all fields except $v$ and $z$  appear with the standard Klein-Gordon Hamiltonian, while $v$ and $z$ have also the extra term $p_zv-p_vz$. Moreover, the masses of $z, x_2,x_3$ depend non-trivially on the deformation parameter $\eta$ and reduce to $1$ only in the undeformed limit.

In principle, the extra term $p_zv-p_vz$ can be removed by a $\tau$-dependent rotation of the $v$ and $z$ fields (see e.g.~\cite{Borsato:2023oru}). 
In the undeformed case ($\eta\to 0$), this is clearly advantageous because all masses become $m_\mu=1$, and one restores a global $SO(2)$ symmetry  that rotates $v$ and $z$. The only consequence of implementing the $\tau$-dependent rotation, then, is to eliminate  $p_zv-p_vz$, so that one arrives to a standard Klein-Gordon Hamiltonian. In the deformed case, however, the $SO(2)$ symmetry is broken since $m_v\neq m_z$. Implementing a $\tau$-dependent rotation to remove $p_zv-p_vz$ would introduce explicit $\tau$ dependence in the Hamiltonian. To avoid this complication, we  prefer to retain   the quadratic Hamiltonian density as written above.\footnote{Let us remark, nevertheless, that a time-dependent Hamiltonian can still be used in principle,  and its quantisation should be equivalent to the approach taken here.}

Before proceeding, let us make a few remarks on an interesting observation  related to Jordanian deformations. In this case, the deformation parameter $\eta$ can be considered unphysical in a certain sense, as it can be fixed to any non-zero value by   exploiting   target-space coordinate transformations in  the deformed background. For example, in~\eqref{eq:G-B-Sch} we can fix $\eta=1$ by rescaling
\begin{equation}
    Z\to \sqrt{\eta}\ Z,\qquad
    x_a\to \sqrt{\eta}\ x_a,\qquad
    V\to \eta\ V. 
\end{equation}
Notice that this transformation is applicable as long as $\eta\neq 0$, which means that after the above rescaling we would lose the interpretation of the background as a continuous deformation of $AdS_5\times S^5$. The possibility of absorbing $\eta$ is related to the interpretation of the Jordanian deformation as a trivial deformation of non-abelian T-duality~\cite{Hoare:2016wsk,Borsato:2016pas,Borsato:2017qsx}.
Notably, under this rescaling, $V$ transforms but $T$ is unaffected. However, when introducing the preferred coordinate system  used for the light-cone gauge-fixing, these coordinates  mix, see e.g.~\eqref{eq:coord-trans-lin}. As a result,  the transformation used to rescale $\eta$ is not compatible with the gauge-fixing employed in this work. Indeed, it is not clear how to fix $\eta=1$ in the gauge-fixed model. In addition, while it would have been possible to set $\eta=1$ in~\eqref{eq:G-B-Sch} and  fix a light-cone gauge from thereon, we would have lost the connection to the undeformed limit to benchmark  results to. Nonetheless, we verified that  the conclusions of our findings in section \ref{s:qH3} would have remained the same  when fixing $\eta=1$ from the outset.

\vspace{12pt}

\noindent \textbf{{The cubic Hamiltonian}} ---
At the next order, one  finds a non-trivial cubic Hamiltonian
\begin{equation}
    \mathcal H_3=\frac{\eta}{\sqrt{1-\eta^2}}\left(v'p_{x_a}x_a-(z+p_v)x_ax'_a+\left(z'-\frac{2\eta}{\sqrt{1-\eta^2}}(z+p_v)\right)x_ax_a\right)-\frac{4\eta^2}{1-\eta^2}z(v^2+p_vz).
\end{equation}
Notice that the cubic Hamiltonian vanishes in the undeformed limit. In fact, the previous target-space coordinate redefinition \eqref{eq:coord-trans-quadcub} has been chosen to ensure this. When sending $\eta\to 0$ one must indeed recover the usual light-cone gauge-fixed Hamiltonian of $AdS_5\times S^5$, albeit written in an inequivalent light-cone gauge as explained in~\cite{Borsato:2023oru}. A non-trivial cubic Hamiltonian is therefore not expected when $\eta\to 0$, and thus it must be possible to cancel possible cubic contributions by field redefinitions and by  dropping total derivative terms. Equivalently, in the $\eta\to 0$ limit there are no cubic processes in the S-matrix, and thus even if one had cubic terms in the Hamiltonian, they would  contribute only trivially to the scattering processes once the external particles are set on-shell.

In principle, one could try to eliminate the cubic Hamiltonian in the deformed case as well. For example, one may use field redefinitions  $x^\mu\to x^\mu +\mathcal O(x^2)$ where transverse fields are shifted by expressions that are quadratic in the fields themselves. These transformations preserve the quadratic Hamiltonian, but modify the cubic and higher-order terms.  Another option is to perform more complicated canonical transformations. 
Importantly, all of these  transformations would not affect  the S-matrix. Therefore,  to keep the discussion simple, we will retain the Hamiltonians ${\cal H}_2$ and ${\cal H}_3$ as written above and proceed to verify whether  cubic scattering processes arise.

\section{Quantisation} \label{s:quant-ham}
The quadratic Hamiltonian density $\mathcal{H}_2$ \eqref{eq:H2} is non-standard. In the undeformed limit $\eta\to 0$ it can be understood as
\begin{equation}
    \mathcal H_2|_{\eta=0} = \mathcal H_2^{KG} + \mathcal Q,
\end{equation}
where $\mathcal H_2^{KG}$ is the Hamiltonian density for eight Klein-Gordon fields with mass $1$, and $\mathcal Q$ can be interpreted as the charge density for the $SO(2)$ symmetry rotating the fields $v$ and $z$. As discussed earlier, this extra term can be reabsorbed by doing a $\tau$-dependent rotation of $v$ and $z$. Alternatively, since the quantum Hamiltonian and the $SO(2)$ charge are mutually diagonalisable, one can directly work with the above Hamiltonian and proceed to quantise it~\cite{Borsato:2023oru}. Using this approach, the creation and annihilation operators introduced for the fields $v$ and $z$ are found to  necessarily mix, 
leading to eigenstates of the Hamiltonian with dispersion relation $\omega_p^q=q+\sqrt{1+p^2}$, where $p$ is the momentum of the excitation and $q$ its charge (with possible values $q=\pm1$ in this case). In the following, we will explain how to generalise this story to the deformed case.

\subsection{Quadratic quantum Hamiltonian}
We want to quantise the fields to obtain a quadratic Hamiltonian $H_2 = \int d\sigma \mathcal H_2$ of the form
\begin{equation} \label{eq:H2-canonical}
    H_2 = \int dp \sum_{I}\omega^{(I)}_p\ a^\dagger_{(I)p}a^{(I)}_{p},
\end{equation}
where the index $I$ runs over all  possible excitations and $\omega^{(I)}_p$ are their corresponding dispersion relations. While   the quantisation of the Klein-Gordon Hamiltonian is straightforward, the non-trivial part lies in the quantisation of  the $v$ and $z$ fields. We find that 
the quadratic Hamiltonian can be diagonalised as in \eqref{eq:H2-canonical} if we perform the following Fourier transformations for the fields\footnote{The expressions for the corresponding conjugate momenta can be derived from the Hamilton equations $\dot x^\mu = \{H_2,x^\mu\}$. For the $v$ and $z$ fields, the relation to the conjugate momenta is non-standard because the Hamiltonian is not of Klein-Gordon form, and one should use $p_v=\dot v+z, p_z=\dot z-v$.}
\begin{equation} \label{eq:fourier-v-z}
    \begin{aligned}
        v(\tau,\sigma) &=\frac{1}{\sqrt{2\pi}}\int dp \sum_{\lambda=+,-}\alpha_{(\lambda)}(p)\left[a^{(\lambda)}(p)e^{-i(\omega^{(\lambda)}_p\tau-p\sigma)}+a^\dagger_{(\lambda)}(p)e^{i(\omega^{(\lambda)}_p\tau-p\sigma)}\right],\\
        z(\tau,\sigma) &=\frac{1}{\sqrt{2\pi}}\int dp\sum_{\lambda=+,-}\ i\beta_{(\lambda)}(p)\left[a^{(\lambda)}(p)e^{-i(\omega^{(\lambda)}_p\tau-p\sigma)}-a^\dagger_{(\lambda)}(p)e^{i(\omega^{(\lambda)}_p\tau-p\sigma)}\right],\\
        x_a(\tau,\sigma) &=\frac{1}{\sqrt{2\pi}}\int \frac{dp}{\sqrt{2\omega^{(x_a)}_p}}\left[a^{(x_a)}(p)e^{-i(\omega^{(x_a)}_p\tau-p\sigma)}+a^\dagger_{(x_a)}(p)e^{i(\omega^{(x_a)}_p\tau-p\sigma)}\right],\\
          y_i(\tau,\sigma) &=\frac{1}{\sqrt{2\pi}}\int \frac{dp}{\sqrt{2\omega^{(s)}_p}}\left[a^{(i)}(p)e^{-i(\omega^{(s)}_p\tau-p\sigma)}+a^\dagger_{(i)}(p)e^{i(\omega^{(s)}_p\tau-p\sigma)}\right],\\
    \end{aligned}
\end{equation}
where the frequencies are
\begin{equation} \label{eq:freqs-allorders}
    \omega^{(\pm)}_p=\sqrt{\left(\pm 1+\sqrt{\frac{1}{\left(1-\eta ^2\right)^2}+p ^2}\right)^2-\frac{\eta ^4}{\left(1-\eta ^2\right)^2}},\qquad
    \omega^{(x_a)}_p=\sqrt{\frac{1+\eta^2}{1-\eta^2}+p^2},\qquad
    \omega^{(s)}_p=\sqrt{1+p^2},
\end{equation}
corresponding to the sets of creation and annihilation operators $(a^\dagger_{(\lambda)},a^{(\lambda)}),(a^\dagger_{(x_a)},a^{(x_a)}),(a^\dagger_{(i)},a^{(i)})$ respectively. We note that at vanishing light-cone momentum $p=0$ the frequency $\omega^{(-)}_p$ vanishes 
\begin{equation}
    \omega^{(-)}_{p=0}=0,
\end{equation} 
and thus the $(-)$ excitations are gapless. %which one could attribute to the existence of the non-compact isometry in the original $v$ direction in target space. 
The real functions appearing in the quantisation of $v$ and $z$ are
\begin{equation}
    \alpha_{(\pm)}(p)=\sqrt{\frac{\omega^{(\pm)}_p\left[\left(\omega^{(\mp)}_p\right)^2-p^2\right]}{2p^2\left[\left(\omega^{(\mp)}_p\right)^2-\left(\omega^{(\pm)}_p\right)^2\right]}},\qquad
     \beta_{(\pm)}(p)=\pm\sqrt{\frac{\left(\omega^{(\pm)}_p\right)^2-p^2}{2\omega^{(\pm)}_p\left[\left(\omega^{(\pm)}_p\right)^2-\left(\omega^{(\mp)}_p\right)^2\right]}}.
\end{equation}
This  Fourier transformation was chosen to ensure also canonical commutation relations for creation and annihilation operators, i.e.
\begin{equation}
    [a_p^{(I)} , a^\dag_{(J)p'}] = \delta^I_J \delta(p-p') , \qquad [a^\dag_{(I)p}, a^\dag_{(J)p'}] =  0 , \qquad[a_p^{(I)} , a_{p'}^{(J)}] = 0 ,
\end{equation}
as a consequence of canonical commutation relations for the phase space fields.

Before proceeding in using this transformation to quantise the cubic Hamiltonian, let us note that around $\eta=0$, and for a fixed momentum $p$, the  expressions expand as 
\begin{equation} \label{eq:disp-rels-eta-2}
    \begin{aligned}
        \omega^{(\pm)}_p= \left(\pm 1 +\sqrt{1+p^2} \right) + \frac{\eta^2}{\sqrt{1+p^2}} + {\cal O}(\eta^4), \qquad
        \omega^{(x_a)}_p= \sqrt{1+p^2}  + \frac{\eta^2}{\sqrt{1+p^2}} + {\cal O}(\eta^4), \qquad
        \omega^{(s)}_p=\sqrt{1+p^2},
    \end{aligned}
\end{equation}
and
\begin{equation}
\begin{aligned}
    \alpha_{(\pm)}(p) &= \frac{1}{2 \sqrt[4]{1+p^2}}  \left( 1+ \frac{1 \mp (1+p^2) \sqrt{1+p^2}}{2p^2 (1+p^2)} \eta^2 \right) + {\cal O}(\eta^4), \\
    \beta_{(\pm)}(p) &=  \pm \frac{1}{2 \sqrt[4]{1+p^2}}  \left(   1 -  \frac{ 1 + 2p^2 \mp (1+p^2) \sqrt{1+p^2}}{2p^2 (1+p^2)} \eta^2 \right) + {\cal O}(\eta^4) .
\end{aligned}
\end{equation}
In the strict undeformed limit, they indeed coincide with an oscillator expansion of fields with shifted dispersion relations  \cite{Borsato:2023oru}. Note that to the next order the dispersion relations of all the $AdS$ fields coincide for fixed $p$, up to the constant shifts $\pm 1$.

\subsection{Cubic quantum Hamiltonian} \label{s:qH3}
So far, we have  quantised only the free part of the Hamiltonian, $H_2$, which was straightforward because $H_2$ is quadratic in the fields. The interaction part of the Hamiltonian, defined as $V = H-H_2$,  is used to compute the scattering matrix $\mathbb S$. In particular, the $\mathbb T$-matrix,  defined via $\mathbb S=1-i \mathbb T$, can be expressed at tree-level as
\begin{equation}
    \mathbb T = \int_{-\infty}^{+\infty} d\tau\ V(\tau)+\ldots.
\end{equation}
At the lowest order in the field expansion, 
we have $V\sim H_3$, so  one could expect e.g.~cubic processes of the types
\begin{equation}
    I + J \to K\qquad \text{or} \qquad I \to J + K,
\end{equation}
where $I,J,K$ represent three generic  particle types from $AdS$. 
To determine whether these processes are indeed present,  
we explicitly compute here the contribution of $H_3$ to $\mathbb T$ when rewriting the fields in terms of the creation and annihilation operators (a summary of our results is presented in table \ref{table:t-matrix-elements-summ}). After normal ordering, in general the cubic part of $\mathbb T$  takes the following form
\begin{equation}
\begin{alignedat}{3}
    \mathbb T_3 ={}& \int d\tau d\sigma\ \mathcal H_3  && \\
        ={}& \int dp_1 dp_2 dp_3 
        &&\Bigg[\delta (p_1+p_2+p_3) \sum_{I,J,K} \delta (\omega^{(I)}_1 + \omega^{(J)}_2 + \omega^{(K)}_3) \\
        & &&\times \left(\mathbf{T}^{(I)(J)(K)}(p_1,p_2,p_3) a_{(K)}^\dag (p_3) a_{(J)}^\dag (p_2) a_{(I)}^\dag (p_1) 
        + \mathbf{T}_{(I)(J)(K)}(p_1,p_2,p_3) a^{(K)} (p_3) a^{(J)} (p_2) a^{(I)} (p_1) \right) \\ 
        & &&+ \bigg[ \delta (p_1-p_2-p_3) \sum_{I,J,K} \delta (\omega^{(I)}_1 - \omega^{(J)}_2 - \omega^{(K)}_3)  \\
        & &&\times \left(\mathbf{T}^{(J)(K)}_{(I)}(p_1,p_2,p_3) a_{(K)}^\dag (p_3) a_{(J)}^\dag (p_2) a^{(I)} (p_1)  
        +\mathbf{T}_{(J)(K)}^{(I)}(p_1,p_2,p_3) a_{(I)}^\dag (p_1) a^{(K)} (p_3) a^{(J)} (p_2) \right)\\
        & &&\quad+ (p_1 \leftrightarrow p_3) +(p_1 \leftrightarrow p_2)  \bigg]\Bigg] .
\end{alignedat}
\end{equation}
Here, the delta functions enforce the conservation of momentum and energy, originating from the integration over $\sigma$ and $\tau$, respectively. As before, indices $I,J,K\in \{(\pm), (x_a), (s)\}$ denote the different   particle types, which appear in the creation and annihilation operators, in the coefficients $\mathbf T^{(I)(J)(K)}$, etc., and in the dispersion relations $\omega^{(I)}_p$.
It is important to distinguish  between the coefficients $\mathbf T^{(I)(J)(K)}$, etc., (identified directly from rewriting the Hamiltonian density $\mathcal{H}_3$ in terms of oscillators) and the elements of the $\mathbb T$-matrix, denoted as  $\mathbb T^{(I)(J)(K)}$, etc., which are obtained after integrating the delta functions. Schematically, these are related to $\mathbb T_3$ as
\begin{equation}
\begin{aligned}
    \mathbb T_3 &= \int dp\ 
   \sum_{I,J,K}  \left(\mathbb{T}^{(I)(J)(K)}(p) a_{(K)}^\dag a_{(J)}^\dag a_{(I)}^\dag  
        + \mathbb{T}_{(I)(J)(K)}(p) a^{(K)} a^{(J)}  a^{(I)} \right.\\
        &\qquad\qquad\qquad\left.+\mathbb{T}^{(J)(K)}_{(I)}(p) a_{(K)}^\dag  a_{(J)}^\dag  a^{(I)}   
        +\mathbb{T}_{(J)(K)}^{(I)}(p) a_{(I)}^\dag  a^{(K)}  a^{(J)}  \right),
\end{aligned}
\end{equation}
where the creation and annihilation operators are assumed to have a momentum dependence that is allowed by the conservation of energy and momentum. See~\eqref{eq:T3-formula} for the relation between $\mathbf T^{(I)(J)(K)}$ and $\mathbb T^{(I)(J)(K)}$.

To begin, from the cubic Hamiltonian $H_3$ one can identify the following non-vanishing coefficients
\begin{equation}
\begin{aligned}
    \mathbf{T}^{(\pm)(\pm)(\pm)},\quad\mathbf{T}^{(\pm)(x_a)(x_a)},\quad
    \mathbf{T}_{(\pm)}^{(\pm)(\pm)},\quad\mathbf{T}_{(\pm)}^{(x_a)(x_a)},\quad\mathbf{T}_{(x_a)}^{(\pm)(x_a)},\\
     \mathbf{T}_{(\pm)(\pm)(\pm)},\quad\mathbf{T}_{(\pm)(x_a)(x_a)},\quad
    \mathbf{T}^{(\pm)}_{(\pm)(\pm)},\quad\mathbf{T}^{(\pm)}_{(x_a)(x_a)},\quad\mathbf{T}^{(x_a)}_{(\pm)(x_a)}.
    \end{aligned}
\end{equation}
Here, the signs are uncorrelated, but the index $a=2,3$ is fixed.\footnote{Recall that there is a residual $SO(2)$ symmetry rotating the $(x_2,x_3)$ fields.}
These coefficients can contribute  to the S-matrix for physical particles if they satisfy the following two conditions: \textit{(i)} momentum and energy conservation can be solved for real momenta, and \textit{(ii)}  the coefficients remain non-vanishing when going on-shell. \\

Let us first consider the case where three particles are annihilated into (or created out of) the vacuum. 
Here it is useful to observe that all dispersion relations are non-negative, and among them $\omega^{(-)}_p$ is the only  one that can vanish, which happens when $p=0$. Consequently, of these processes, the only  kinematically allowed cases are $(-)+(-)+(-)\rightarrow 0$ 
and  $0\rightarrow (-)+(-)+(-)$ (the reversed process). They can occur if    energy-momentum conservation holds 
\begin{equation}
    p_1 + p_2 + p_3 = 0 , \qquad \omega^{(I)}_1 + \omega^{(J)}_2 + \omega^{(K)}_3 = 0 ,
\end{equation}
which has the only solution $p_1=p_2=p_3=0$. The corresponding elements  $\mathbb{T}^{(-)(-)(-)}$ and $\mathbb{T}_{(-)(-)(-)}$ appear to diverge due to collinear divergences (e.g.~$p_1\to -p_3$) and IR divergences (i.e.~when  momenta go to zero).\footnote{Explicitly, after imposing the momentum conservation as $p_2=-p_1-p_3$ and up to terms that are zero in the limit of soft momentum, 
the expression becomes proportional to  $\sqrt{\frac{1}{p_3}+\frac{1}{p_1}}$, $\sqrt{\frac{p_1}{p_3
   (p_1+p_3)}}$, and $\sqrt{\frac{p_3}{p_1 (p_1+p_3)}}$.} We attribute this feature to the gaplessness of the particles of type $(-)$.\\

Let us now examine   processes where one particle decays into two, or conversely, where two particles merge into one. For these cases, energy-momentum conservation imposes
\begin{equation}
    p_1 = p_2 + p_3 , \qquad \omega^{(I)}_1 = \omega^{(J)}_2 + \omega^{(K)}_3 .
\end{equation}
It is important to note that the dispersion relations \eqref{eq:freqs-allorders} are generally non-relativistic and  can be quite intricate.
Explicit  checks are therefore needed to determine whether solutions for the energy-momentum conservation exist.

First, let us focus on processes that only involve $(+),(-)$ excitations. Solving the condition for  energy conservation at finite $\eta$ can be challenging due to the complexity of the dispersion relations. One may do a numerical analysis, which shows that  the processes are either  kinematically forbidden ($\mathbb{T}^{(+)(-)}_{(-)},\mathbb{T}^{(+)(+)}_{(-)},\mathbb{T}^{(+)(+)}_{(+)}$, and reversed) or have vanishing amplitude ($\mathbb{T}^{(-)(-)}_{(+)},\mathbb{T}^{(-)(-)}_{(-)},\mathbb{T}^{(+)(-)}_{(+)}$, and reversed).

A second family of processes  are those with elements $\mathbb{T}^{(I)(-)}_{(I)}$ with $I=+,-,x$ (and the corresponding reversed processes), where a particle of type $(I)$ emits (or absorbs) a ``soft'' particle of type $(-)$, i.e.~with  momentum  exactly zero.\footnote{For $I=+,-$ these cases have already been considered in the previous paragraph, but it is useful to have a discussion here for general $I$.} It is clear that this is indeed a possible solution for the conservation of energy and momentum, as this is a direct consequence of the gaplessness of the $(-)$-type dispersion relation. 
We verified analytically (at generic finite values of $\eta$) that all matrix elements $\mathbb{T}^{(I)(-)}_{(I)}$ vanish for a soft  $(-)$ particle. This requires taking a careful limit as the momentum of the soft particle approaches  zero, to avoid indeterminate expressions.

Finally, we  consider the remaining cubic processes, namely $\mathbb{T}_{(\pm)}^{(x_a)(x_a)},\mathbb{T}_{(x_a)}^{(\pm)(x_a)}$ (and their reversed versions) where particularly in the latter case the $(-)$ particle is not soft. 
Due to the complexity of solving  energy conservation at finite $\eta$,  we carry out this analysis in an expansion around small $\eta$. First, in the undeformed limit $\eta\to 0$, the only kinematically allowed processes  are
\begin{equation}\label{eq:cubic-x}
   (+) \rightarrow (x_a) + (x_a)    , \qquad (x_a) \rightarrow  (-) + (x_a) ,
\end{equation}
with solutions
\begin{equation} \label{eq:cubic-x-solem}
    p_2=0,\ p_3=p_1,\qquad\qquad \text{or}\qquad\qquad p_2=p_1,\ p_3=0.
\end{equation}
In the first process $(+) \rightarrow (x_a) + (x_a)$, the two solutions \eqref{eq:cubic-x-solem} are equivalent because the $(x_a)$ particles are indistinguishable. In the second process $(x_a) \rightarrow  (-) + (x_a)$, the solution with $p_2=0$ yields a soft gapless $(-)$ particle, which we already analysed in the previous paragraph. Therefore, for both cases, we can focus  on the solution   $p_2=p_1,\ p_3=0$ at $\eta=0$, where  the  particle with vanishing momentum  is always gapped. 
Turning on $\eta$,  energy-momentum conservation now deforms this solution to
\begin{equation}\label{eq:sol-p3}
  p_2 = p_1-p_3=p_1- \frac{\sqrt{1+p_1^2}}{p_1} \eta^2 + {\cal O}(\eta^4)   ,\qquad p_3 = \eta^2\frac{\sqrt{1+p_1^2}}{p_1} + {\cal O}(\eta^4) .
\end{equation}
Note that the $p_1\rightarrow 0$ limit is ill-defined. 
For simplicity, we now assume that $p\equiv p_1>0$ is positive. To compute the T-matrix elements, we will  integrate over $p_2$ by imposing momentum conservation ($p_2=p_1-p_3$) and over $p_3$ by imposing energy conservation \eqref{eq:sol-p3}. In the latter case, one needs to take into account also the Jacobian. In particular, if the energy conservation is imposed with the delta function $\delta(f_{(I)}^{(J)(K)}(p_3))$, so that 
\begin{equation}
     f_{(I)}^{(J)(K)}(p_3)=0,\qquad \text{with}\quad  f_{(I)}^{(J)(K)}(p_3) = \omega^{(I)}(p) - \omega^{(J)}(p-p_3) - \omega^{(K)}(p_3),
\end{equation}
then  the Jacobian is
\begin{equation}
    J_{(I)}^{(J)(K)}= \left\vert\partial_{p_3}f_{(I)}^{(J)(K)}(p_3)\right\vert_{\text{\tiny on-shell}},
\end{equation}
where ``on-shell''  means evaluating the expression on the solution for $p_3$ given in~\eqref{eq:sol-p3}.
For the processes in~\eqref{eq:cubic-x}, the Jacobian evaluates to
\begin{equation}
     J_{(+)}^{(x)(x)} =  J_{(x)}^{(-)(x)} =\frac{p}{\sqrt{1+p^2}}+\frac{\eta ^2 \left(-1-3 p^2-p^4-\sqrt{1+p^2}\right)}{p
   \left(1+p^2\right)^{3/2}}+\mathcal O(\eta^4).
\end{equation}
The  corresponding T-matrix elements are then given by
\begin{equation}\label{eq:T3-formula}
    \mathbb T_{(I)}^{(J)(K)} = \left. \frac{(2\pi)^2}{J_{(I)}^{(J)(K)}}\mathbf T_{(I)}^{(J)(K)}\right|_{\text{\tiny on-shell}} .
\end{equation}
Explicitly, we find\footnote{Although we do not write  the reversed processes, they can be found by complex conjugation. Note, furthermore, that when writing this result, we took into account that the coefficients $\mathbf T_{(I)}^{(J)(K)}$   extracted from $\mathcal H_3$ are of ${\cal O}(\eta)$, so that we can write down the $\mathbb T_{(I)}^{(J)(K)}$ up to ${\cal O}(\eta^3)$ included, since we solved  energy-momentum conservation to ${\cal O}(\eta^2)$.}
\begin{equation}
    \begin{aligned}
   \mathbb{T}^{(x)(x)}_{(+)}  &= - \frac{\sqrt{2\pi}}{6} \eta - i\frac{\sqrt{2\pi} (1-\sqrt{1+p^2})}{6p} \eta^2 - \frac{\sqrt{2\pi}((1+2p^2)(1+\sqrt{1+p^2})+2)}{12 p^2 \sqrt{1+p^2}} \eta^3 + { \cal O}(\eta^4)\\ 
     \mathbb{T}^{(-)(x)}_{(x)} &= \frac{\sqrt{2\pi}}{6} \eta - i\frac{\sqrt{2\pi}  \left(1+\sqrt{1+p^2} \right)}{6 p} \eta^2 + \frac{\sqrt{2\pi} \left((1+2p^2) \left(-1+\sqrt{1+p^2}\right)\right)}{12p^2 \sqrt{1+p^2}} \eta^3 + {\cal O}(\eta^4) .
    \end{aligned}
\end{equation}
These cubic S-matrix elements are therefore non-vanishing.  We ran also numerical checks to confirm that the above expressions accurately describe the amplitudes when $\eta$ is fixed to a small numerical value (so that one can safely discard terms of ${\cal O}(\eta^4)$ or higher). Furthermore,  one can also solve the complicated condition of energy-momentum conservation for fixed numerical values of $p_2$ and $p_3$ and general finite values of $\eta\in (0,1)$, and we verified that then the elements $\mathbb{T}^{(x)(x)}_{(+)}$ and $\mathbb{T}^{(-)(x)}_{(x)}$ are indeed non-zero.

Before concluding, let us briefly comment on the level-matching condition. As in the undeformed case (see e.g.~\cite{Arutyunov:2004yx,Arutyunov:2009ga}),  fixing the light-cone gauge on the plane and solving the Virasoro constraints leads to the requirement that  the total worldsheet momentum density should vanish  for physical states. However,  to describe  $N$-particle scattering in a model with factorised scattering,  the level-matching condition must be relaxed for the $2\to 2$ body S-matrix, as individual factors in the factorisation do not necessarily need to satisfy it. In other words, while the total momentum of the $N$-particle state vanishes, $\sum_{i=1}^Np_i=0$, the sum of the momenta of any subset of two particles from the $N$ particles is not necessarily zero. For this reason,  we do not impose the level-matching condition in our analysis. However, if we did  enforce it, we find that there is no single process of the type $1\to 2$ nor $2\to 1$. For instance, in $(+)\to (x) + (x)$, energy-momentum conservation with $p_1=0$ can be satisfied only for imaginary values of $p_2,p_3$. For $(x)\to (-) + (x)$, the conservation of energy-momentum with $p_1=0$ requires $p_2=0$, implying that  the $(-)$ particle is soft.  As  mentioned previously, in that case the corresponding scattering element vanishes. Although it is reassuring that these processes disappear under the level-matching condition, in general we want to relax level-matching to allow for more general scattering dynamics. Particle production in intermediate processes would in particular be incompatible with a possible factorisation of scattering and, therefore, S-matrix integrability.\\

\begin{table}
\centering
\begin{tabular}{lll}
\toprule
\textbf{$\mathbb{T}$-matrix element} & \textbf{Result} & \textbf{Verification method} \\
\toprule
$\mathbb{T}^{(-)(-)(-)}$ & diverges & analytical \\
\midrule
$\mathbb{T}^{(+)(\pm)(\pm)}$            & forbidden & analytical \\
$\mathbb{T}^{(+)(\pm)}_{(-)}$           & forbidden & numerical \\
$\mathbb{T}^{(+)(+)}_{(+)}$             & forbidden & numerical \\
$\mathbb{T}^{(x)(x)}_{(-)}$              & forbidden & expansion in $\eta$ and numerical \\
$\mathbb{T}^{(+)(x)}_{(x)}$              & forbidden & expansion in $\eta$ and numerical \\
\midrule
$\mathbb{T}^{(-)(-)}_{(\pm)}$            & zero      & numerical \\
$\mathbb{T}^{(+)(-)}_{(+)}$            & zero      & numerical \\
$\mathbb{T}^{(I)(-)}_{(I)}$ with soft $(-)$  & zero      & analytical \\
\midrule
$\mathbb{T}^{(x)(x)}_{(+)}$             & non-zero  & expansion in $\eta$ and numerical \\
$\mathbb{T}^{(-)(x)}_{(x)}$ with non-soft $(-)$ & non-zero  & expansion in $\eta$ and numerical \\
\bottomrule
\end{tabular}
\caption{Summary of $\mathbb{T}$-matrix elements for creation processes grouped by outcome and verification method. The reversed processes yield identical results. The index $(I)$ labels all AdS excitations $I=+,-,x$.}
\label{table:t-matrix-elements-summ}
\end{table}

We summarise our results  in table \ref{table:t-matrix-elements-summ}.
In conclusion, 
the Jordanian model under study  exhibits tree-level processes where the number of particles is not conserved. We will discuss the consequences of this result in the next section.

\section{Final discussion and outlook} \label{s:discussion}

We considered strings on a  Jordanian-deformed $AdS_5\times S^5$ background, and focused on the worldsheet scattering of its bosonic sigma-model. 
After fixing light-cone gauge, using a pointlike classical solution
that reduces to the BMN vacuum when  the deformation parameter vanishes,
we found surprising features in the gauge-fixed theory. 
Specifically, there are tree-level scattering processes that do not conserve the number of particles, since cubic vertices are non-vanishing on-shell. We included a \textsf{Mathematica} notebook to the \textsf{arXiv} submission of this paper, where we summarise the main calculations  needed to arrive at this result. Interestingly, however, when we impose level-matching the $1\to 2$ and $2\to 1$ processes that we observed do not contribute.  Nevertheless, when relaxing the level-matching condition, the  presence of non-trivial cubic processes is incompatible with the usual notion of integrable scattering, for which the conservation of the number of particles  is an essential requirement.\footnote{There exist, however, exceptions: see for example~\cite{Kozlowski:2016too} where $1\to 2$ and $2\to 1$ processes are still compatible with integrability. In that case, the cubic process represent the fusion of non-relativistic Galilean particles of masses $m_1$ and $m_2$ into a particle of mass $m_1+m_2$ (and vice-versa). There is an infinite number of species of particles, with all non-negative integer masses allowed, and the conservation of the total number of particles is replaced by the conservation of the total mass.}
At the same time, the sigma-model under study is integrable in the sense that  it admits a Lax connection  prior to gauge-fixing. While these two notions of integrability are distinct, according to the usual lore  they are traditionally regarded as closely connected.
In the following, we begin with some general considerations and proceed to discuss the implications of our findings,  as well as potential explanations for  the tension between  Lax integrability and the observed worldsheet scattering dynamics.

\vspace{12pt}

\noindent \textbf{{The cubic Hamiltonian}} ---
After obtaining a cubic Hamiltonian, and before computing  the scattering elements explicitly, one might wonder whether  $\mathcal{H}_3$ can be eliminated by certain operations. One could, for example, implement  redefinitions of the transverse fields of the form $x^\mu \to x^\mu + \mathcal O(x^2)$. 
These are shifts that are quadratic in the transverse fields themselves, so that they leave the quadratic Hamiltonian $\mathcal H_2$ invariant, whilst generating new cubic terms  from $\mathcal H_2$ that could in principle cancel contributions from  $\mathcal H_3$. 
Other options include more complicated canonical transformations, or dropping total derivatives in $\mathcal H_3$, or terms that vanish on the equations of motion.
However, these operations can only yield equivalent forms of $\mathcal H_3$ and can not modify the scattering processes derived from the Hamiltonian. The ``experimental fact'' that for this model it is not possible to cancel $\mathcal H_3$ by these operations is therefore in agreement with our finding that there are non-trivial cubic scattering processes.

\vspace{12pt}

\noindent \textbf{{The sigma model}} ---
Because we are interested in tree-level scattering processes, we dealt only with the classical bosonic action in this paper. This means that  our input data was  the metric and NSNS-flux of the Jordanian background, whilst we  ignored the contribution from the dilaton, $F_{3}$-, and $F_{5}$-flux, which complete this  background to type IIB supergravity. 
Indeed, the $F_{3}$-, and $F_{5}$-flux couple necessarily to fermionic degrees of freedom, and the Fradkin-Tseytlin term, which couples the dilaton to the worldsheet Ricci curvature, contributes only at the next order in $\alpha'$ (or inverse string tension $g$). There is then no reason to consider these extra fields to compute the tree-level bosonic scattering. Regarding the case of the Fradkin-Tseytlin term, one may insist and wonder if it should actually be included, and if it may help to eliminate unwanted cubic processes, on the argument that
different orders in $g^{-1}$ may mix when
 implementing the rescaling in $g^{-1}$ of the worldsheet fields in the decompactification limit.
A related comment is that,
to impose light-cone gauge,
one needs to solve the Virasoro constraints. These correspond to the equations of motion of the worldsheet metric, and they receive a contribution from the dilaton term at the next order in $\alpha'$. Nevertheless, not only do  these considerations appear to be far-fetched for our purposes, they are also not relevant for the model that we consider. In fact,  the background of~\cite{Kawaguchi:2014fca}  has a constant dilaton, so that the Fradkin-Tseytlin term is a total derivative and thus  does not contribute.

\vspace{12pt}

\noindent \textbf{{Light-cone gauge and S-matrix}} ---
For  sigma models in light-cone gauge, we know that the S-matrix depends on the specific gauge choice \cite{Borsato:2023oru}.
One may therefore wonder if the result of having on-shell cubic processes would change if, while still expanding around the same classical solution, the light-cone gauge is fixed in a different way. Concretely, this means that one identifies  the ``transverse'' and  ``longitudinal'' fields differently, where the latter are the fields that are gauge-fixed. This can be achieved by implementing diffeomorphisms that mix $x^\pm$ with $x^\mu$ \emph{before} gauge-fixing, and then setting  $x^+=\tau,p_-=1$ for the new variables. Alternative gauges of this kind were considered in~\cite{Borsato:2023oru}. However, the consequences on the S-matrix were analysed therein, albeit under certain assumptions, and they would not be able to  eliminate cubic terms in the scattering matrix.

\vspace{12pt}

\noindent \textbf{{Local higher-spin charges}} ---
As  mentioned, the violation of particle number conservation in a 1+1 dimensional field theory is incompatible with the usual definition of S-matrix integrability. Only certain special models enjoy integrable scattering, and this is typically understood as the consequence of an infinite number of symmetries in involution.
The typical argument  starts from the assumption that there is an infinite number of commuting  higher-spin \emph{local} charges~\cite{Parke:1980ki}. These can be simultaneously conserved  under the scattering process only if the number of particles is conserved, and if their momenta are simply reshuffled. In addition, their existence also implies factorisation of scattering.
For the type of sigma-models that we consider, commuting  higher-spin local charges cannot  be typically extracted from the monodromy matrix of the flat Lax connection. 
Nevertheless, following the construction of~\cite{Lacroix:2017isl},\footnote{This work generalises previous constructions, in particular those for the PCM and symmetric space sigma-model, of~\cite{Evans:1999mj,Evans:2000qx}.} 
they can be built from the  Lax itself, and therefore they exist, 
at least at the classical level, 
also for the integrable Jordanian deformation considered here.
One might then believe that they should place constraints on the  scattering processes to ensure integrable scattering.

Despite the above considerations, the argument that commuting higher-spin local charges imply integrable scattering should be treated with care. In fact,  \emph{any} sigma-model, even those that do not admit a flat Lax connection, possesses a tower of commuting higher-spin local charges. Indeed, scale invariance of the classical 1+1 dimensional theory implies that the conservation of the stress-energy tensor reads  $\partial_\pm T_{\mp\mp}=0$ and, therefore, powers of $T_{\pm\pm}$ can be used to construct the  higher-spin local charges as $ \int d\sigma (T_{\pm\pm})^n$.\footnote{Since we are only interested in tree-level scattering, we do not  worry about possible quantum anomalies for these charges.} The naive expectation that  this tower of charges could imply integrable scattering, at least at tree-level, is clearly too strong. 
A first important point is that the arguments of~\cite{Parke:1980ki} are valid for  field theories with \emph{massive} particles. For string sigma-models at tree-level, as considered here, the massive spectrum appears only after fixing light-cone gauge. Furthermore, the construction of~\cite{Evans:1999mj,Evans:2000qx,Lacroix:2017isl} is valid for sigma-models that are not coupled to a dynamical worldsheet metric. In our case, we could go to that setup  if we fixed conformal gauge, but that would not be useful to achieve the description of the scattering problem that is interesting for us, as in that case already the undeformed worldsheet model will not be massive.
Given our choice to fix light-cone gauge,  it is legitimate to wonder if the infinite tower of commuting local  charges are genuinly present. Under the assumption that they can be constructed for a dynamical worldsheet metric, they would be useful for the integrability argument only if  
their action on states of the gauge-fixed theory is not trivial, otherwise, the  usual argument to claim S-matrix integrability would be inapplicable.\footnote{In fact,  this trivialisation of  charges in the light-cone gauge is precisely what occurs for the infinite tower of local charges derived from the stress-energy tensor, as they vanish identically once  the Virasoro constraints $T_{\alpha\beta}=0$ are imposed.}
Additionally, fixing the light-cone gauge explicitly breaks  worldsheet Lorentz invariance, and this has two important consequences. First, without Lorentz symmetry one cannot  assign a spin to the would-be conserved charges. Second, this  means that our case does not fall into the  assumptions of~\cite{Parke:1980ki}, which rely on Lorentz symmetry to   conclude that  the infinite tower of commuting charges implies the absence of particle production and factorisation of scattering.

\vspace{12pt}

\noindent \textbf{{Gapless dispersion relation}} ---
Even in a scenario where  an infinite tower of commuting local charges survive the light-cone gauge-fixing, our model has an additional complication due to the presence of the gapless excitation with dispersion relation $\omega^{(-)}_p$,\footnote{Let us note here that its presence can be traced back to the existence of the (non-compact) shift isometry of the deformed background \eqref{eq:G-B-Sch} in the coordinate $V$, which persists after employing the alternative light-cone gauge.} which may undermine the usual arguments of integrable scattering. As previously noted, a key assumption in the arguments is that the particles involved are  massive~\cite{Parke:1980ki}, i.e.~their dispersion relations have a gap at zero  momentum. This allows the construction of  wave packets  that can be spatially separated. However, when gapless particles are in the game, there could be a loophole in the reasoning. Works that discuss the presence  of gapless particles  and the   tension  that this causes between Lax integrability and the absence of tree-level particle production are, e.g.,~\cite{Nappi:1979ig,Hoare:2018jim,Georgiou:2024jlz}.
However, it  is important to note that  our situation differs from~\cite{Nappi:1979ig,Hoare:2018jim,Georgiou:2024jlz}, and that the issues raised therein are not directly relevant for our discussion. In those works, tree-level  particle production arises from the need to regulate  internal massless propagators in  Feynman diagrams, which would otherwise lead to indeterminate expressions.  Typical choices of regularisation appear to be incompatible with absence of particle production \cite{Hoare:2018jim}. It was recently realised, however, that an ad hoc prescription leading to compatibility is possible in certain cases, such as the $SU(2)$ Principal Chiral Model~\cite{Georgiou:2024jlz}. In our case, the issue of regularisation is not a concern,  because we  find cubic vertices on the nose from the classical Hamiltonian. Nonetheless, it is true that the applicability of the usual arguments for integrable scattering remains questionable due to the presence of the gapless particle with  dispersion relation $\omega^{(-)}_p$. 
However, while we do have the cubic process $(x_a) \rightarrow  (-) + (x_a) $ involving this  particle, we also observe the non-trivial process $(+) \rightarrow (x_a) + (x_a)  $, which exclusively involves gapped particles. For this reason, we believe that  the gaplessness of the $(-)$ excitation  is unlikely to  be the primary or only explanation for the observed particle production.

\vspace{12pt}

\noindent \textbf{{The asymptotic particle spectrum}} ---
Given the existence of processes where particles can decay into pairs or, conversely, pairs of particles can merge into one, one may ask whether compatibility with integrability could be restored by reducing the asymptotic particle spectrum. In other words, one may want to declare that only a restricted subset of particles should be treated as ``fundamental'', while the others are interpreted as ``composite'' particles. This is in fact what happens in the worldsheet descriptions of the integrable backgrounds $AdS_4\times \mathbb{ CP}^3$~\cite{Zarembo:2009au} and  $AdS_3\times S^3\times S^3\times S^1$~\cite{Sundin:2012gc}. In those cases, certain ``heavy modes''  are considered composite rather than fundamental because, when including quantum corrections, the naive pole in their two-point function becomes a branch cut. Therefore, these heavy modes can  be consistently removed from the asymptotic spectrum and accounted for in the Bethe ansatz via stacks of Bethe roots. In our model, however,  the decay processes that we observe do not  seem to allow room for this interpretation. There is simply no hierarchy that we can assign to particles to consistently interpret the  decay processes as ``composite'' $\to$ ``fundamental'' $+$  ``fundamental''.

\vspace{12pt}

\noindent \textbf{{Alternative classical solutions}} ---
Leaving aside the question of why  particle production occurs, it may happen that the issue will disappear if the light-cone gauge is fixed using a different classical solution. The classical solution used here might simply represent a ``bad'' vacuum, while a more suitable vacuum would lead to a potential that forbids  decay processes and  possibly has only massive particles. 
For example, one can easily generalise the classical pointlike solution used in this paper by allowing  for a rotation of the pointlike string in the $(x_2,x_3)$ plane. However, while a more general family of pointlike solutions may be possible (which is something that we did not explore),  it is worth pointing out that setting $x_2=x_3=0$ is a consistent truncation of the deformed theory. In the undeformed limit, this corresponds to truncating to an $AdS_3$ geometry parametrised by the coordinates $T,V,Z$, which can be easily verified by examining the embedding coordinates given in Eq.~(A.2) of~\cite{Borsato:2022drc}.
Turning on the deformation parameter $\eta$, the consistent truncation then corresponds to a Jordanian deformation of $AdS_3$. In this setup, and considering the residual shift isometries of $T,V$, a suitable pointlike string ansatz would be $T=\alpha \tau, V=\beta\tau, Z=f(\tau)$, with $\alpha, \beta$ arbitrary real constants and $f$ an arbitrary  function to be fixed. One will then find that the equations of motion   enforce  $f$ to be constant, so that we effectively return to the  classical solution considered in this paper. In particular, one would still have the gapless excitation $(-)$, making the usual interpretation of integrability challenging. Nevertheless, although it is not useful to the current setup, it is interesting to note that in the $AdS_3$ truncation we would not have  $1\to 2$ and $2\to 1$ processes. We have not explored the light-cone gauge fixing around classical solutions of the deformed theory that are not pointlike.

\vspace{12pt}

\noindent \textbf{{Drinfel'd twisted S-matrix}} ---
The original motivation for this project was to test the general expectation that  a Homogeneous Yang-Baxter deformation should modify the worldsheet S-matrix  by a Drinfel'd twist. This is in fact what happens in the case of the $\beta$-deformation and more general ``diagonal'' TsT deformations of $AdS_5\times S^5$~\cite{vanTongeren:2021jhh}, see also~\cite{Beisert:2005if,Ahn:2010ws}. Drinfel'd twists also play a key role in formulating   the corresponding deformations of the dual gauge theory~\cite{vanTongeren:2015uha,Araujo:2017jkb,Guica:2017mtd,Meier:2023kzt,Meier:2023lku}. However, an important distinction arises for Yang-Baxter deformations beyond the diagonal TsT class, as they break 
the light-cone isometries normally used to fix light-cone gauge. 
To consistently fix the gauge in such deformed setups, an alternative light-cone gauge-fixing has to be employed already in the undeformed limit, as discussed in~\cite{Borsato:2023oru}. As also remarked earlier, changing the gauge generally modifies the S-matrix. Therefore, the object expected to be Drinfel'd twisted is not the well-known worldsheet S-matrix of $AdS_5\times S^5$ \cite{Beisert:2005tm,Arutyunov:2006yd}, but rather its gauge-transformed version.
 What we find in this paper is that in the undeformed limit we recover the picture expected from~\cite{Borsato:2023oru}, but the connection to the Drinfel'd twist  becomes unclear because the scattering is not factorised.
Furthermore, it is also not obvious how to interpret the action of the Drinfel'd twist  itself on the asymptotic states of the undeformed  theory in the alternative gauge. 
Specifically, one could consider the  Drinfel'd twist expected for the Jordanian model \cite{10.1007/BFb0101176,Ogievetsky1994}. While the symmetry generators required for the twisting belong to the  symmetry algebra that survives after implementing the alternative light-cone gauge in the \emph{undeformed} model, their action on the transverse fields is non-linear. This non-linearity thus further complicates the interpretation of the Drinfel'd twist operation on the S-matrix.

\vspace{12pt}

\noindent \textbf{{The twisted open string picture}} ---
An interesting direction  to try to understand the worldsheet scattering problem is to use the map relating the Jordanian deformation to an undeformed sigma-model with twisted boundary conditions. At the classical level, it is known that the Yang-Baxter deformation of a sigma-model can be undone at the cost of imposing boundary conditions that are not periodic~\cite{Frolov:2005dj,Vicedo:2015pna,vanTongeren:2018vpb,Borsato:2021fuy}. The explicit on-shell map between the Yang-Baxter deformed model and the corresponding undeformed but twisted model was constructed in~\cite{Borsato:2021fuy} and subsequently used in~\cite{Borsato:2022drc,Driezen:2024mcn} to extract the semi-classical spectrum for the Jordanian deformation studied here. One could try to fix a gauge using the classical solution for the twisted open string that corresponds to the classical pointlike solution considered in this paper, and then analyse the scattering of the physical worldsheet excitations. Given that the on-shell transformation between the deformed and the twisted models is non-local, one could hope that   integrable scattering is restored in the twisted model. We hope to come back to this question in the future. 

\vspace{12pt}

\noindent \textbf{{Final remarks}} ---
\noindent Understanding how to reconcile  integrability with this Jordanian deformation (and potentially other Homogeneous Yang-Baxter deformations) remains an open and intriguing question. In particular, it would be interesting to investigate whether the particle production observed here is specific to the Jordanian case or if it also arises  in other Yang-Baxter deformations that break the BMN light-cone isometries. For this, addressing the simplest possible ``non-diagonal'' TsT deformations could provide crucial insights. Equally intriguing is  the challenge that this result poses to the conventional expectation regarding the relationship between Lax  and S-matrix integrability.
It is not necessarily evident that the Lax integrability of a (string) sigma-model prior to gauge-fixing should imply integrable scattering in the gauge-fixed theory. If this does not happen, we hope that some points in the above discussion may offer clues on what the mechanisms behind the breakdown of S-matrix integrability could be. We believe that trying to comprehend this issue would be an important step in improving our common understanding of what it  means for a string sigma-model to be integrable.

\vspace{12pt}

\section*{Acknowledgements}
We thank Sergey Frolov, Ben Hoare, Sylvain Lacroix, Nat Levine, J.~Luis Miramontes, Davide Polvara, Fiona Seibold,  Kostas Sfetsos, and Arkady Tseytlin for useful discussions. We  thank Sergey Frolov and Ben Hoare for reading the draft and sharing their comments.
We also gratefully acknowledge the participants and  organisers of  the workshops \textit{Integrability in Low Supersymmetry Theories} (Filicudi 2023), \textit{Integrable Sigma-Models} (Z\"urich 2024),  and \textit{Exact approaches to low-supersymmetry AdS/CFT} (SCGP 2024), as well as support from the Simons Center for Geometry and Physics, Stony Brook University, at which some of the research for this paper was performed.
The work of R.B.~was supported by the grants RYC2021-032371-I (funded by MCIN/AEI/10.13039/501100011033 and by the European Union ``NextGenerationEU''/PRTR), 2023-PG083 (with reference code ED431F 2023/19 funded by Xunta de Galicia), and PID2023-152148NB-I00 (funded by AEI-Spain). R.B.~also acknowledges the Mar\'ia de Maeztu grant CEX2023-001318-M (funded by MICIU/AEI /10.13039/501100011033),  Xunta de Galicia (CIGUS Network of Research Centres), and the European Union. 
The work of S.D.~is supported by the Swiss National
Science Foundation through the SPF fellowship TMPFP2$\_224600$ and the NCCR SwissMAP.

\bibliographystyle{nb}
\bibliography{biblio}{}

%bibliography generated by nb.bst v1.01 (C) 2003-2010 Niklas Beisert
\begin{thebibliography}{10}
\ifx\href\asklfhas\newcommand{\href}[2]{#2}\fi
\ifx\arxivref\asklfhas\newcommand{\arxivref}[2]{\href{http://arxiv.org/abs/#1}{#2}}\fi
\ifx\doiref\asklfhas\newcommand{\doiref}[2]{\href{http://dx.doi.org/#1}{#2}}\fi
\raggedright
\small
\parskip 0pt

\bibitem{Beisert:2010jr}
N.~Beisert et~al.,
\textit{``{Review of AdS/CFT Integrability: An Overview}''},
\textsf{\doiref{10.1007/s11005-011-0529-2}{Lett.~Math.~Phys.~99,~3~(2012)}},
\texttt{\arxivref{1012.3982}{arxiv:1012.3982}}.
%%CITATION = ARXIV:1012.3982;%%

\bibitem{Klimcik:2008eq}
C.~Klimcik,
\textit{``{On integrability of the Yang-Baxter sigma-model}''},
\textsf{\doiref{10.1063/1.3116242}{J.Math.Phys.~50,~043508~(2009)}},
\texttt{\arxivref{0802.3518}{arxiv:0802.3518}}.
%%CITATION = ARXIV:0802.3518;%%

\bibitem{Delduc:2013fga}
F.~Delduc, M.~Magro and B.~Vicedo,
\textit{``{On classical $q$-deformations of integrable sigma-models}''},
\textsf{\doiref{10.1007/JHEP11(2013)192}{JHEP~1311,~192~(2013)}},
\texttt{\arxivref{1308.3581}{arxiv:1308.3581}}.
%%CITATION = ARXIV:1308.3581;%%

\bibitem{Delduc:2013qra}
F.~Delduc, M.~Magro and B.~Vicedo,
\textit{``{An integrable deformation of the AdS$_5 \times$S$^5$ superstring
  action}''},
\textsf{\doiref{10.1103/PhysRevLett.112.051601}{Phys.Rev.Lett.~112,~051601~(2014)}},
\texttt{\arxivref{1309.5850}{arxiv:1309.5850}}.
%%CITATION = ARXIV:1309.5850;%%

\bibitem{Kawaguchi:2014qwa}
I.~Kawaguchi, T.~Matsumoto and K.~Yoshida,
\textit{``{Jordanian deformations of the $AdS_5 x S^5$ superstring}''},
\textsf{\doiref{10.1007/JHEP04(2014)153}{JHEP~1404,~153~(2014)}},
\texttt{\arxivref{1401.4855}{arxiv:1401.4855}}.
%%CITATION = ARXIV:1401.4855;%%

\bibitem{vanTongeren:2015soa}
S.~J.~van~Tongeren,
\textit{``{On classical Yang-Baxter based deformations of the AdS$_{5}$ ×
  S$^{5}$ superstring}''},
\textsf{\doiref{10.1007/JHEP06(2015)048}{JHEP~1506,~048~(2015)}},
\texttt{\arxivref{1504.05516}{arxiv:1504.05516}}.
%%CITATION = ARXIV:1504.05516;%%

\bibitem{vanTongeren:2015uha}
S.~J.~van~Tongeren,
\textit{``{Yang–Baxter deformations, AdS/CFT, and twist-noncommutative gauge
  theory}''},
\textsf{\doiref{10.1016/j.nuclphysb.2016.01.012}{Nucl.~Phys.~B904,~148~(2016)}},
\texttt{\arxivref{1506.01023}{arxiv:1506.01023}}.
%%CITATION = ARXIV:1506.01023;%%

\bibitem{drinfeld1983constant}
V.~G.~Drinfeld,
\textit{``Constant quasiclassical solutions of the Yang--Baxter quantum
  equation''},
in: \textit{``Doklady Akademii Nauk''},
531--535p.

\bibitem{Giaquinto:1994jx}
A.~Giaquinto and J.~J.~Zhang,
\textit{``{Bialgebra actions, twists, and universal deformation formulas}''},
\textsf{\doiref{10.1016/S0022-4049(97)00041-8}{J.~Pure~Appl.~Algebra~128,~133~(1998)}},
\texttt{\arxivref{hep-th/9411140}{hep-th/9411140}}.

\bibitem{Kulish2009}
P.~Kulish,
\textit{``Twist Deformations of Quantum Integrable Spin Chains''},
in: \textit{``Noncommutative Spacetimes: Symmetries in Noncommutative Geometry
  and Field Theory''},
Springer Berlin Heidelberg (2009),
Berlin, Heidelberg,
167--190p,
\href{https://doi.org/10.1007/978-3-540-89793-4\_9}{\texttt{https://doi.org/10.1007/978-3-540-89793-4\_9}}.

\bibitem{Osten:2016dvf}
D.~Osten and S.~J.~van~Tongeren,
\textit{``{Abelian Yang–Baxter deformations and TsT transformations}''},
\textsf{\doiref{10.1016/j.nuclphysb.2016.12.007}{Nucl.~Phys.~B915,~184~(2017)}},
\texttt{\arxivref{1608.08504}{arxiv:1608.08504}}.
%%CITATION = ARXIV:1608.08504;%%

\bibitem{Reshetikhin:1990ep}
N.~Reshetikhin,
\textit{``{Multiparameter quantum groups and twisted quasitriangular Hopf
  algebras}''},
\textsf{\doiref{10.1007/BF00626530}{Lett.~Math.~Phys.~20,~331~(1990)}}.

\bibitem{Beisert:2005if}
N.~Beisert and R.~Roiban,
\textit{``{Beauty and the twist: The Bethe ansatz for twisted N=4 SYM}''},
\textsf{\doiref{10.1088/1126-6708/2005/08/039}{JHEP~0508,~039~(2005)}},
\texttt{\arxivref{hep-th/0505187}{hep-th/0505187}}.
%%CITATION = HEP-TH/0505187;%%

\bibitem{Ahn:2010ws}
C.~Ahn, Z.~Bajnok, D.~Bombardelli and R.~I.~Nepomechie,
\textit{``{Twisted Bethe equations from a twisted S-matrix}''},
\textsf{\doiref{10.1007/JHEP02(2011)027}{JHEP~1102,~027~(2011)}},
\texttt{\arxivref{1010.3229}{arxiv:1010.3229}}.

\bibitem{deLeeuw:2012hp}
M.~de~Leeuw and S.~J.~van~Tongeren,
\textit{``{The spectral problem for strings on twisted $AdS_5 \times S^5$}''},
\textsf{\doiref{10.1016/j.nuclphysb.2012.03.004}{Nucl.~Phys.~B860,~339~(2012)}},
\texttt{\arxivref{1201.1451}{arxiv:1201.1451}}.
%%CITATION = ARXIV:1201.1451;%%

\bibitem{Kazakov:2018ugh}
V.~Kazakov,
\textit{``{Quantum Spectral Curve of $\gamma$-twisted ${\cal N}=4$ SYM theory
  and fishnet CFT}''},
\texttt{\arxivref{1802.02160}{arxiv:1802.02160}},
in: \textit{``Ludwig Faddeev Memorial Volume''},
ed.: M.-L.~Ge, A.~J.~Niemi, K.~K.~Phua and L.~A.~Takhtajan,
293--342p.

\bibitem{vanTongeren:2021jhh}
S.~J.~van~Tongeren and Y.~Zimmermann,
\textit{``{Do Drinfeld twists of $AdS_5 \times S^5$ survive light-cone
  quantization?}''},
\textsf{\doiref{10.21468/SciPostPhysCore.5.2.028}{SciPost~Phys.~Core~5,~028~(2022)}},
\texttt{\arxivref{2112.10279}{arxiv:2112.10279}}.

\bibitem{Guica:2017mtd}
M.~Guica, F.~Levkovich-Maslyuk and K.~Zarembo,
\textit{``{Integrability in dipole-deformed ${\mathcal{N}=4}$ super
  Yang–Mills}''},
\textsf{\doiref{10.1088/1751-8121/aa8491}{J.~Phys.~A50,~394001~(2017)}},
\texttt{\arxivref{1706.07957}{arxiv:1706.07957}}.
%%CITATION = ARXIV:1706.07957;%%

\bibitem{Vicedo:2015pna}
B.~Vicedo,
\textit{``{Deformed integrable \ensuremath{\sigma}-models, classical R-matrices
  and classical exchange algebra on Drinfel\textquoteright{}d doubles}''},
\textsf{\doiref{10.1088/1751-8113/48/35/355203}{J.~Phys.~A~48,~355203~(2015)}},
\texttt{\arxivref{1504.06303}{arxiv:1504.06303}}.

\bibitem{Frolov:2005dj}
S.~Frolov,
\textit{``{Lax pair for strings in Lunin-Maldacena background}''},
\textsf{\doiref{10.1088/1126-6708/2005/05/069}{JHEP~0505,~069~(2005)}},
\texttt{\arxivref{hep-th/0503201}{hep-th/0503201}}.
%%CITATION = HEP-TH/0503201;%%

\bibitem{vanTongeren:2018vpb}
S.~J.~Van~Tongeren,
\textit{``{On Yang--Baxter models, twist operators, and boundary
  conditions}''},
\textsf{\doiref{10.1088/1751-8121/aac8eb}{J.~Phys.~A~51,~305401~(2018)}},
\texttt{\arxivref{1804.05680}{arxiv:1804.05680}}.

\bibitem{Borsato:2021fuy}
R.~Borsato, S.~Driezen and J.~L.~Miramontes,
\textit{``{Homogeneous Yang-Baxter deformations as undeformed yet twisted
  models}''},
\textsf{\doiref{10.1007/JHEP04(2022)053}{JHEP~2204,~053~(2022)}},
\texttt{\arxivref{2112.12025}{arxiv:2112.12025}}.

\bibitem{Borsato:2016ose}
R.~Borsato and L.~Wulff,
\textit{``{Target space supergeometry of $\eta$ and $\lambda$-deformed
  strings}''},
\textsf{\doiref{10.1007/JHEP10(2016)045}{JHEP~1610,~045~(2016)}},
\texttt{\arxivref{1608.03570}{arxiv:1608.03570}}.
%%CITATION = ARXIV:1608.03570;%%

\bibitem{Borsato:2022ubq}
R.~Borsato and S.~Driezen,
\textit{``{All Jordanian deformations of the $AdS_5 \times S^5$
  superstring}''},
\textsf{\doiref{10.21468/SciPostPhys.14.6.160}{SciPost~Phys.~14,~160~(2023)}},
\texttt{\arxivref{2212.11269}{arxiv:2212.11269}}.

\bibitem{vanTongeren:2016eeb}
S.~J.~van~Tongeren,
\textit{``{Almost abelian twists and AdS/CFT}''},
\textsf{\doiref{10.1016/j.physletb.2016.12.002}{Phys.~Lett.~B765,~344~(2017)}},
\texttt{\arxivref{1610.05677}{arxiv:1610.05677}}.
%%CITATION = ARXIV:1610.05677;%%

\bibitem{Araujo:2017jkb}
T.~Araujo, I.~Bakhmatov, E.~O.~Colg\'{a}in, J.~Sakamoto, M.~M.~Sheikh-Jabbari
  and K.~Yoshida,
\textit{``{Yang-Baxter $\sigma$-models, conformal twists, and noncommutative
  Yang-Mills theory}''},
\textsf{\doiref{10.1103/PhysRevD.95.105006}{Phys.~Rev.~D95,~105006~(2017)}},
\texttt{\arxivref{1702.02861}{arxiv:1702.02861}}.
%%CITATION = ARXIV:1702.02861;%%

\bibitem{Leigh:1995ep}
R.~G.~Leigh and M.~J.~Strassler,
\textit{``{Exactly marginal operators and duality in four-dimensional N=1
  supersymmetric gauge theory}''},
\textsf{\doiref{10.1016/0550-3213(95)00261-P}{Nucl.~Phys.~B~447,~95~(1995)}},
\texttt{\arxivref{hep-th/9503121}{hep-th/9503121}}.

\bibitem{Lunin:2005jy}
O.~Lunin and J.~M.~Maldacena,
\textit{``{Deforming field theories with $U(1) \times U(1)$ global symmetry and
  their gravity duals}''},
\textsf{\doiref{10.1088/1126-6708/2005/05/033}{JHEP~0505,~033~(2005)}},
\texttt{\arxivref{hep-th/0502086}{hep-th/0502086}}.
%%CITATION = HEP-TH/0502086;%%

\bibitem{Seiberg:1999vs}
N.~Seiberg and E.~Witten,
\textit{``{String theory and noncommutative geometry}''},
\textsf{\doiref{10.1088/1126-6708/1999/09/032}{JHEP~9909,~032~(1999)}},
\texttt{\arxivref{hep-th/9908142}{hep-th/9908142}}.
%%CITATION = HEP-TH/9908142;%%

\bibitem{Bergman:2000cw}
A.~Bergman and O.~J.~Ganor,
\textit{``{Dipoles, twists and noncommutative gauge theory}''},
\textsf{\doiref{10.1088/1126-6708/2000/10/018}{JHEP~0010,~018~(2000)}},
\texttt{\arxivref{hep-th/0008030}{hep-th/0008030}}.

\bibitem{Meier:2023kzt}
T.~Meier and S.~J.~van~Tongeren,
\textit{``{Quadratic Twist-Noncommutative Gauge Theory}''},
\textsf{\doiref{10.1103/PhysRevLett.131.121603}{Phys.~Rev.~Lett.~131,~121603~(2023)}},
\texttt{\arxivref{2301.08757}{arxiv:2301.08757}}.

\bibitem{Meier:2023lku}
T.~Meier and S.~J.~van~Tongeren,
\textit{``{Gauge theory on twist-noncommutative spaces}''},
\textsf{\doiref{10.1007/JHEP12(2023)045}{JHEP~2312,~045~(2023)}},
\texttt{\arxivref{2305.15470}{arxiv:2305.15470}}.

\bibitem{Arutyunov:2004yx}
G.~Arutyunov and S.~Frolov,
\textit{``{Integrable Hamiltonian for classical strings on AdS$_5
  \times$S$^5$}''},
\textsf{\doiref{10.1088/1126-6708/2005/02/059}{JHEP~0502,~059~(2005)}},
\texttt{\arxivref{hep-th/0411089}{hep-th/0411089}}.
%%CITATION = HEP-TH/0411089;%%

\bibitem{Arutyunov:2005hd}
G.~Arutyunov and S.~Frolov,
\textit{``{Uniform light-cone gauge for strings in AdS$_5 \times$S$^5$: Solving
  $\mathfrak{su}(1|1)$ sector}''},
\textsf{\doiref{10.1088/1126-6708/2006/01/055}{JHEP~0601,~055~(2006)}},
\texttt{\arxivref{hep-th/0510208}{hep-th/0510208}}.
%%CITATION = HEP-TH/0510208;%%

\bibitem{Arutyunov:2006gs}
G.~Arutyunov, S.~Frolov and M.~Zamaklar,
\textit{``{Finite-size Effects from Giant Magnons}''},
\textsf{\doiref{10.1016/j.nuclphysb.2006.12.026}{Nucl.Phys.~B778,~1~(2007)}},
\texttt{\arxivref{hep-th/0606126}{hep-th/0606126}}.
%%CITATION = HEP-TH/0606126;%%

\bibitem{Arutyunov:2009ga}
G.~Arutyunov and S.~Frolov,
\textit{``{Foundations of the AdS$_5 \times$S$^5$ Superstring. Part I}''},
\textsf{\doiref{10.1088/1751-8113/42/25/254003}{J.Phys.~A42,~254003~(2009)}},
\texttt{\arxivref{0901.4937}{arxiv:0901.4937}}.
%%CITATION = ARXIV:0901.4937;%%

\bibitem{Borsato:2023oru}
R.~Borsato, S.~Driezen, B.~Hoare, A.~L.~Retore and F.~K.~Seibold,
\textit{``{Inequivalent light-cone gauge-fixings of strings on
  AdSn\texttimes{}Sn backgrounds}''},
\textsf{\doiref{10.1103/PhysRevD.109.106023}{Phys.~Rev.~D~109,~106023~(2024)}},
\texttt{\arxivref{2312.17056}{arxiv:2312.17056}}.

\bibitem{Frolov:2019xzi}
S.~Frolov,
\textit{``{$T{\overline T}$, $\widetilde JJ$, $JT$ and $\widetilde JT$
  deformations}''},
\textsf{\doiref{10.1088/1751-8121/ab581b}{J.~Phys.~A~53,~025401~(2020)}},
\texttt{\arxivref{1907.12117}{arxiv:1907.12117}}.

\bibitem{Hoare:2023zti}
B.~Hoare, A.~L.~Retore and F.~K.~Seibold,
\textit{``{Elliptic deformations of the~$\mathsf{AdS}_3 \times \mathsf{S}^3
  \times \mathsf{T}^4$ string}''},
\textsf{\doiref{10.1007/JHEP04(2024)042}{JHEP~2404,~042~(2024)}},
\texttt{\arxivref{2312.14031}{arxiv:2312.14031}}.

\bibitem{Ouyang:2017yko}
H.~Ouyang,
\textit{``{Semiclassical spectrum for BMN string in $Sch_5\times S^5$}''},
\textsf{\doiref{10.1007/JHEP12(2017)126}{JHEP~1712,~126~(2017)}},
\texttt{\arxivref{1709.06844}{arxiv:1709.06844}}.

\bibitem{Borsato:2022drc}
R.~Borsato, S.~Driezen, J.~M.~Nieto~Garc\'\i{}a and L.~Wyss,
\textit{``{Semiclassical spectrum of a Jordanian deformation of
  AdS5\texttimes{}S5}''},
\textsf{\doiref{10.1103/PhysRevD.106.066015}{Phys.~Rev.~D~106,~066015~(2022)}},
\texttt{\arxivref{2207.14748}{arxiv:2207.14748}}.

\bibitem{Driezen:2024mcn}
S.~Driezen and N.~Kamath,
\textit{``{Regularising spectral curves for homogeneous Yang-Baxter
  strings}''},
\textsf{\doiref{10.1016/j.physletb.2024.138971}{Phys.~Lett.~B~857,~138971~(2024)}},
\texttt{\arxivref{2406.09811}{arxiv:2406.09811}}.

\bibitem{vanTongeren:2019dlq}
S.~J.~van~Tongeren,
\textit{``{Unimodular jordanian deformations of integrable superstrings}''},
\textsf{\doiref{10.21468/SciPostPhys.7.1.011}{SciPost~Phys.~7,~011~(2019)}},
\texttt{\arxivref{1904.08892}{arxiv:1904.08892}}.

\bibitem{Kawaguchi:2014fca}
I.~Kawaguchi, T.~Matsumoto and K.~Yoshida,
\textit{``{A Jordanian deformation of AdS space in type IIB supergravity}''},
\textsf{\doiref{10.1007/JHEP06(2014)146}{JHEP~1406,~146~(2014)}},
\texttt{\arxivref{1402.6147}{arxiv:1402.6147}}.
%%CITATION = ARXIV:1402.6147;%%

\bibitem{Blau:2009gd}
M.~Blau, J.~Hartong and B.~Rollier,
\textit{``{Geometry of Schrodinger Space-Times, Global Coordinates, and
  Harmonic Trapping}''},
\textsf{\doiref{10.1088/1126-6708/2009/07/027}{JHEP~0907,~027~(2009)}},
\texttt{\arxivref{0904.3304}{arxiv:0904.3304}}.

\bibitem{Matsumoto:2014ubv}
T.~Matsumoto and K.~Yoshida,
\textit{``{Yang-Baxter deformations and string dualities}''},
\textsf{\doiref{10.1007/JHEP03(2015)137}{JHEP~1503,~137~(2015)}},
\texttt{\arxivref{1412.3658}{arxiv:1412.3658}}.

\bibitem{Kruczenski:2004kw}
M.~Kruczenski, A.~Ryzhov and A.~A.~Tseytlin,
\textit{``{Large spin limit of AdS$_5 \times$S$^5$ string theory and low-energy
  expansion of ferromagnetic spin chains}''},
\textsf{\doiref{10.1016/j.nuclphysb.2004.05.028}{Nucl.Phys.~B692,~3~(2004)}},
\texttt{\arxivref{hep-th/0403120}{hep-th/0403120}}.
%%CITATION = HEP-TH/0403120;%%

\bibitem{Kruczenski:2004cn}
M.~Kruczenski and A.~A.~Tseytlin,
\textit{``{Semiclassical relativistic strings in S$^5$ and long coherent
  operators in N=4 SYM theory}''},
\textsf{\doiref{10.1088/1126-6708/2004/09/038}{JHEP~0409,~038~(2004)}},
\texttt{\arxivref{hep-th/0406189}{hep-th/0406189}}.
%%CITATION = HEP-TH/0406189;%%

\bibitem{Hoare:2016wsk}
B.~Hoare and A.~A.~Tseytlin,
\textit{``{Homogeneous Yang-Baxter deformations as non-abelian duals of the
  AdS$_5$ sigma-model}''},
\textsf{\doiref{10.1088/1751-8113/49/49/494001}{J.~Phys.~A49,~494001~(2016)}},
\texttt{\arxivref{1609.02550}{arxiv:1609.02550}}.
%%CITATION = ARXIV:1609.02550;%%

\bibitem{Borsato:2016pas}
R.~Borsato and L.~Wulff,
\textit{``{Integrable Deformations of $T$-Dual $\sigma$ Models}''},
\textsf{\doiref{10.1103/PhysRevLett.117.251602}{Phys.~Rev.~Lett.~117,~251602~(2016)}},
\texttt{\arxivref{1609.09834}{arxiv:1609.09834}}.
%%CITATION = ARXIV:1609.09834;%%

\bibitem{Borsato:2017qsx}
R.~Borsato and L.~Wulff,
\textit{``{On non-abelian T-duality and deformations of supercoset string
  sigma-models}''},
\textsf{\doiref{10.1007/JHEP10(2017)024}{JHEP~1710,~024~(2017)}},
\texttt{\arxivref{1706.10169}{arxiv:1706.10169}}.
%%CITATION = ARXIV:1706.10169;%%

\bibitem{Kozlowski:2016too}
K.~K.~Kozlowski, E.~Sklyanin and A.~Torrielli,
\textit{``{Quantization of the Kadomtsev\textendash{}Petviashvili equation}''},
\textsf{\doiref{10.1134/S0040577917080074}{Theor.~Math.~Phys.~192,~1162~(2017)}},
\texttt{\arxivref{1607.07685}{arxiv:1607.07685}}.

\bibitem{Parke:1980ki}
S.~J.~Parke,
\textit{``{Absence of Particle Production and Factorization of the $S$ Matrix
  in (1+1)-dimensional Models}''},
\textsf{\doiref{10.1016/0550-3213(80)90196-0}{Nucl.~Phys.~B~174,~166~(1980)}}.

\bibitem{Lacroix:2017isl}
S.~Lacroix, M.~Magro and B.~Vicedo,
\textit{``{Local charges in involution and hierarchies in integrable
  sigma-models}''},
\textsf{\doiref{10.1007/JHEP09(2017)117}{JHEP~1709,~117~(2017)}},
\texttt{\arxivref{1703.01951}{arxiv:1703.01951}}.

\bibitem{Evans:1999mj}
J.~M.~Evans, M.~Hassan, N.~J.~MacKay and A.~J.~Mountain,
\textit{``{Local conserved charges in principal chiral models}''},
\textsf{\doiref{10.1016/S0550-3213(99)00489-7}{Nucl.~Phys.~B~561,~385~(1999)}},
\texttt{\arxivref{hep-th/9902008}{hep-th/9902008}}.

\bibitem{Evans:2000qx}
J.~M.~Evans and A.~J.~Mountain,
\textit{``{Commuting charges and symmetric spaces}''},
\textsf{\doiref{10.1016/S0370-2693(00)00566-9}{Phys.~Lett.~B~483,~290~(2000)}},
\texttt{\arxivref{hep-th/0003264}{hep-th/0003264}}.

\bibitem{Nappi:1979ig}
C.~R.~Nappi,
\textit{``{Some Properties of an Analog of the Nonlinear $\sigma$ Model}''},
\textsf{\doiref{10.1103/PhysRevD.21.418}{Phys.~Rev.~D21,~418~(1980)}}.
%%CITATION = PHRVA,D21,418;%%

\bibitem{Hoare:2018jim}
B.~Hoare, N.~Levine and A.~A.~Tseytlin,
\textit{``{On the massless tree-level S-matrix in 2d sigma models}''},
\textsf{\doiref{10.1088/1751-8121/ab0b79}{J.~Phys.~A~52,~144005~(2019)}},
\texttt{\arxivref{1812.02549}{arxiv:1812.02549}}.

\bibitem{Georgiou:2024jlz}
G.~Georgiou,
\textit{``{The massless S-matrix of integrable $\sigma$-models}''},
\texttt{\arxivref{2408.03673}{arxiv:2408.03673}}.

\bibitem{Zarembo:2009au}
K.~Zarembo,
\textit{``{Worldsheet spectrum in AdS(4)/CFT(3) correspondence}''},
\textsf{\doiref{10.1088/1126-6708/2009/04/135}{JHEP~0904,~135~(2009)}},
\texttt{\arxivref{0903.1747}{arxiv:0903.1747}}.
%%CITATION = ARXIV:0903.1747;%%

\bibitem{Sundin:2012gc}
P.~Sundin and L.~Wulff,
\textit{``{Classical integrability and quantum aspects of the AdS(3) x S(3) x
  S(3) x S(1) superstring}''},
\textsf{\doiref{10.1007/JHEP10(2012)109}{JHEP~1210,~109~(2012)}},
\texttt{\arxivref{1207.5531}{arxiv:1207.5531}}.
%%CITATION = ARXIV:1207.5531;%%

\bibitem{Beisert:2005tm}
N.~Beisert,
\textit{``{The su$(2|2)$ dynamic S-matrix}''},
\textsf{\doiref{10.4310/ATMP.2008.v12.n5.a1}{Adv.Theor.Math.Phys.~12,~945~(2008)}},
\texttt{\arxivref{hep-th/0511082}{hep-th/0511082}}.
%%CITATION = HEP-TH/0511082;%%

\bibitem{Arutyunov:2006yd}
G.~Arutyunov, S.~Frolov and M.~Zamaklar,
\textit{``{The Zamolodchikov-Faddeev algebra for AdS$_5 \times$S$^5$
  superstring}''},
\textsf{\doiref{10.1088/1126-6708/2007/04/002}{JHEP~0704,~002~(2007)}},
\texttt{\arxivref{hep-th/0612229}{hep-th/0612229}}.
%%CITATION = HEP-TH/0612229;%%

\bibitem{10.1007/BFb0101176}
M.~Gerstenhaber, A.~Giaquinto and S.~D.~Schack,
\textit{``Quantum symmetry''},
in: \textit{``Quantum Groups''},
ed.: P.~P.~Kulish,
Springer Berlin Heidelberg (1992),
Berlin, Heidelberg,
9--46p.

\bibitem{Ogievetsky1994}
O.~Ogievetsky,
\textit{``Hopf structures on the Borel subalgebra of $sl(2)$''},
in: \textit{``Proceedings of the Winter School "Geometry and Physics"''},
Circolo Matematico di Palermo (1994),
[185]-199p,
\href{http://eudml.org/doc/221213}{\texttt{http://eudml.org/doc/221213}}.

\end{thebibliography}

\end{document}